\documentstyle[11pt,aaspp4]{article}

\newcommand{\MSUNYR}{\rm M_{\odot}\,yr^{-1}}
\newcommand{\Mdot}{ \dot{M}}
\newcommand{\MSUN}{\rm M_{\odot}}
\newcommand{\RSUN}{\rm R_{\odot}}
\newcommand{\LSUN}{\rm L_{\odot}}
\newcommand{\et}{et al.\ }
\newcommand{\AU}{\rm AU}
\newcommand{\nh}{NH$_3$}
\newcommand{\kms}{\mbox{km s$^{-1}$}}

\slugcomment{Submitted to {\it The Astrophysical Journal}}
\lefthead{G\'{o}mez \& D'Alessio}
\righthead{Line emission from protoplanetary disks}

\begin{document}

\title
{ Resolving Molecular Line Emission from Protoplanetary Disks: 
Observational Prospects for Disks Irradiated by Infalling Envelopes}
\author{Jos\'e F. G\'omez} 
\affil{Laboratorio de Astrof\'\i sica Espacial y F\'\i sica
  Fundamental, INTA\\
  Apartado Correos 50727, E-28080 Madrid, Spain; Electronic
  mail:  jfg@laeff.esa.es}
\and
\author{Paola D'Alessio}
\affil{Instituto de Astronom\'\i a, Universidad Nacional
  Aut\'onoma  de M\'exico\\
  Apartado Postal 70-264, 04510 M\'exico D.F., M\'exico; Electronic mail: 
  dalessio@astroscu.unam.mx}

\begin{abstract}
Molecular line observations that could resolve protoplanetary disks
of $\sim 100$ AU both spatially and kinematically would be 
a useful tool 
to unambiguously identify these disks and to determine 
their kinematical and physical
characteristics.
In this work we model the expected line emission from 
a protoplanetary disk irradiated by an infalling envelope, addressing
the question of its detectability with subarcsecond resolution. 
We adopt a previously determined 
disk model structure that gives a 
continuum spectral energy distribution and a mm intensity spatial
distribution that are  consistent with observational constraints
of HL Tau. 
An analysis of the capability of presently working 
and projected interferometers at mm and submm wavelengths shows that
molecular transitions of moderate opacity 
at these wavelengths (e.g., C$^{17}$O lines)
are good candidates for detecting disk lines at subarcsecond
resolution in the near future. We suggest that, in general, disks of typical
Class I sources will be detectable. Higher line intensities are obtained
for lower inclination angles, larger turbulent velocities, and 
higher temperatures, with less effect from density changes.

The resulting maps show several characteristics that can be tested
observationally. 
A clear asymmetry in the line intensity, with more intense emission in
the disk area farther away from the observer, can be used to compare
the geometrical relationship between disks and outflows. A decrease in
intensity towards the center of the disk is also evident. Finally, 
the emission
peaks in position velocity diagrams trace mid-plane Keplerian
velocities.

\end{abstract}


\section{Introduction}

Circumstellar accretion disks around young stellar objects (YSOs) play
a central role in our understanding of the stellar and planetary
formation processes and related phenomena, which justifies the
important efforts invested in their detection and characterization.

In the currently accepted paradigm, these accretion disks should 
appear naturally from the collapse of a
core of molecular gas with a residual angular momentum. 
Theoretical models
estimate that disks of radius $\sim 100$ AU should form around young
low-mass stars (see, e.g., Terebey, Shu, \& Cassen 1984\markcite{ter84}; 
Morfill 1989\markcite{mor89}).
Gas and dust disks of sizes $\sim 100$ AU could give rise to
planetary systems similar to ours. This is why these disks are usually
referred to as ``protoplanetary disks''.

The extensive search for protoplanetary disks that has taken place
during the past years have provided astronomers with a large body of
compelling evidence about their existence. Most of this evidence is
indirect in the sense that it comes from observations that do not
resolve the emission from the disk or that infer their existence from
different phenomena in the gas and dust (e.g.,
Rodr\'{\i}guez \et 1986\markcite{rod86}; Adams, Shu, \&
Lada 1988\markcite{ada88}; Bertout, Basri, \& Bouvier 1988\markcite{ber88};
Kenyon, Hartmann, \& Hewett 1988\markcite{ken88}; Keene \&
Masson 1990\markcite{kee90}; Carr \et 1993\markcite{car93}; O'Dell, Wen, \&
Hu 1993\markcite{ode93}; Stauffer \et 1994\markcite{sta94}).
A major difficulty to
obtain a resolved image of a protoplanetary disk is that 
subarcsecond resolution is required 
(100 AU subtends 0\farcs 7 at 140 pc, the
distance to the closest known star-disk systems). 

The most unambiguous identification of a disk, and the 
highest level of information about its physical parameters would 
probably be obtained with spectroscopic observations, 
with good enough angular resolution to resolve
them, i.e, subarcsecond resolution. In addition to the morphological
information that continuum observations can provide, 
spectroscopic observations would
yield kinematical information from the gas in the disk, namely the
kinematical signatures of rotation. Spectroscopic
observations will also be a very powerful tool to determine the
physical characteristics (e.g., densities and temperatures) of disks without
requiring heavy modeling. 
If we can combine spectroscopic capabilities with subarcsecond
resolution, we will then obtain 
a conclusive evidence for
a structure to be considered a protoplanetary disk, and very accurate
constraints on its physics. 

Unfortunately, this kind of observations has not been possible so far
with the current instrumentation. Important observational efforts have
been devoted to obtain resolved continuum images of disks or
unresolved spectroscopic data. 
Continuum observations have provided information on sizes and
morphology of disks, as well as the relationship of these structures
with outflow phenomena (e.g.,
Rod\'{\i}guez \et 1992\markcite{rod92}, 1994\markcite{rod94}; Lay \et 
1994\markcite{lay94}; Mundy \et 1996\markcite{mun96}; Wilner, Ho, \&
Rodr\'{\i}guez 1996\markcite{wil96}; 
McCaughream \& O'Dell 1996\markcite{mcc96}; Burrows \et
1996\markcite{bur96}). On the other hand, spectroscopic data have
provided kinematical information, and clues on the physical properties
of disks, specially in the inner regions for the CO overtone emission 
(e.g., Hartmann \& Kenyon 1987a, 1987b\markcite{har87a}\markcite{har87b}; 
Calvet \et 1991\markcite{cal91}; \ Carr \et
1993\markcite{car93}; 
Najita \et 1996\markcite{naj96}; 
Najita et al. 1999\markcite{naj99})

In practice, to obtain resolved spectroscopic data, 
one would probably have to pursue 
molecular line observations of disks using radio
interferometers in the near future. 
First, because these instruments can achieve high angular
resolutions. Second, because lines in disks with temperatures $\sim
100$ K (Beckwith \et 1990\markcite{bec90})
are more likely to be detected in the
radio regime. And third, because the absorption due to molecular
material external to the disk is lower at radio wavelengths.

Of course, the key question is whether it is possible to resolve and
detect disks with sizes on the order of 100 AU, using 
molecular line observations. However, there
are not many instruments that can reach subarcsecond resolution at
present. One of the possibilities, using presently working
instruments, was to observe inversion transitions of ammonia using the
Very Large Array (VLA), whose B configuration have an angular
resolution of $\sim 0\farcs 4$ for ammonia lines at 1.3 cm. This
experiment was carried out 
by G\'{o}mez \et (1993\markcite{gom93})
toward HL Tau and L1551-IRS 5, but
unfortunately it gave only upper limits. Looking into the future, we
may wonder whether the resolving of protoplanetary disks will be
possible with instruments now in project, 
for instance the Millimeter Array (MMA) of
the National Radio Astronomy Observatory, the Submillimeter Array
(SMA) of the Smithsonian Astrophysical Observatory, the Large
Millimeter Array of the Nobeyama Radio Observatory, any possible
upgrade of the present millimeter interferometers, the new VLA
receivers at 7 mm (of which 13 are already working), etc (see Ho
1995\markcite{ho95}). 

To answer the question of the future
detectability of protoplanetary disks, we have developed a model of
their molecular line emission. In particular, 
this paper presents the case of 
an accretion disk irradiated by an infalling envelope, 
constructed to satisfy 
the observational constraints in continuum of HL Tau.
There are
several models of molecular line emission from protoplanetary disks
that have been reported in the literature. These models have been
developed essentially to compare with and to predict results of
observations feasible with present telescopes, i.e., with beams $\geq
1''$ that do not resolve the disks. For instance, Beckwith \&
Sargent (1993\markcite{bec93}), and Omodaka, Kitamura, \&
Kawazoe (1992\markcite{omo92}) 
presented models of
the emission of the rotational transitions of the CO isotopes at 3 mm,
finding a double-peaked line profile, which is typical of rotating
disks. In general, the comparison of these models with the
observations gives good indirect evidence for the existence of these
disks. There are also models for the emission of larger disk-like
structures ($R\geq 1000$ AU) in Keplerian rotation, which also compare
well with observations (e.g., GG Tau: Dutrey, Guilloteau, \&
Simon 1994\markcite{dut94}; GM Aur: Koerner, Sargent, \&
Beckwith 1993\markcite{koe93}).
 Although the sizes
of these structures are about one order of magnitude larger than the
actual accretion disks (which are believed to lie within these larger
structures), these works are very important since they give good
support to the theoretical expectation of the existence of rotating
disks around young stellar objects. However, in some cases the 
interpretation of flattened structures of sizes on the order of 1000
AU as large rotating disks is not straightforward. For instance, 
the kinematics in the $^{13}$CO 
structure mapped in HL Tau (Sargent \& Beckwith 1991\markcite{sar91};
Hayashi, Ohashi, \& Miyama 1993\markcite{hay93}),
could be dominated by an infalling flattened 
envelope (Hartmann \et 1994\markcite{har94}; Hartmann, Calvet \&
Boss 1996\markcite{har96}), 
an infalling 
``pseudodisk'' formed in the presence of a magnetic field (Galli \& 
Shu 1993a,b\markcite{gal93a}\markcite{gal93b}; Hayashi
\et 1993\markcite{hay93}), 
or entrainment in a bipolar outflow (Cabrit \et
1996\markcite{cab96}).

This paper addresses the problem of calculating 
the molecular line emission
from protoplanetary disks (radius $\sim 100$ AU) expected when
observing with an arbitrary angular resolution, especially the
subarcsecond resolution necessary to resolve the disks. In this first
work, we focus on the case of a disk irradiated by an infalling
envelope. This would be the case for embedded sources, in an early
stage of evolution. Apart from the
radial dependence of density and temperature within the disk, we
include, for the first time, a detailed vertical dependence of these
parameters in the calculation of molecular line emission.  In particular, 
we calculate molecular line profiles for the case
of HL Tau, which is probably the disk source most extensively studied in the
literature. The wealth of data from this source allow us to constrain
the density and temperature structure (both radial and vertical) of
its disk, mainly by fitting its continuum emission. From this
structure, we computed the molecular line emission that, therefore, is
consistent with the continuum emission.  
Thus, our calculations 
apply to a realistic source structure, 
that we can assume then as a good example of
star-disk system in an early stage of evolution, still embedded in a
substantial amount of circumstellar material.
From our calculations we can have an estimate about
whether the imaging of molecular line emission from this type of 
protoplanetary  disks
is possible with the telescopes presently available, or we have to
wait for the advent of the next generation of radio interferometers to
achieve such a goal.

The model presented here will be complemented with future
similar studies using models for disks at different stages of
evolution, like, for instance, optically visible T Tauri stars. In
those cases, the properties of the envelope and the physical processes
in the disk will produce a different density and temperature
structure, which will affect line detectability and other
observational properties. Those different models will
require an individualized treatment.

\section{Model for the Disk Structure in HL Tau} 
\label{structure}
HL Tau is a classical T Tauri star with a strong excess emission
at both optical-UV and far IR-radio wavelengths, which is still embedded in
circumstellar material (Stapelfeld et al.\markcite{sta95} 1995).  
The large infrared excess
can be explained by emission from an infalling
dusty envelope (Calvet \et 1994\markcite{cal94}, hereafter CHKW;
Hartmann \et 1996\markcite{har96}, hereafter HCB), 
which also   
 reproduce other observational
features like redshifted C$_2$ optical absorption lines 
(Grasdalen \et 1989\markcite{gra89}), 
near infrared scattered light images 
(Beckwith, Koreski, \& Sargent 1989\markcite{bec89})
and the velocity pattern seen in the spatially-resolved $^{13}$CO map
obtained with 
the Nobeyama Millimeter Array (Hayashi \et 1993\markcite{hay93}).
 However, the envelope surrounding HL Tau cannot account for the
observed continuum flux
at wavelengths
longer than $\sim 1$ mm (CHKW, HCB).

On the other hand, 
a circumstellar disk can have enough column mass to
explain the observed millimeter continuum  emission of HL Tau, if the 
temperature in the outer regions of the disk is sufficiently high 
(Beckwith \et 1990\markcite{bec90}; Beckwith \&
Sargent 1991\markcite{bec91}),
with a dependency with distance to the star
roughly $T\propto r^{-0.5}$.
The interferometric observations of Lay
\et (1994\markcite{lay94}, 1997\markcite{lay97}) at $\lambda=0.65, 0.87$ 
mm with the
single baseline CSO-JCMT, Sargent \& Koerner at $\lambda=1.4 $ mm with
OVRO (see Lay \et 1997\markcite{lay97}), 
Mundy \et (1996\markcite{mun96}) at 2.7 mm 
using BIMA, and 
Wilner, Ho, \& Rodr\'{\i}guez (1996\markcite{wil96}) at 7 mm
using the VLA in its B configuration, 
 indicate that the millimeter continuum emission 
is confined to small scales $R \lesssim  200$ AU, suggesting that
it comes from a disk.

 D'Alessio, Calvet, \& Hartmann (1997\markcite{dal97}, hereafter DCH) 
have developed 
models of steady accretion disks irradiated by optically thick 
infalling envelopes. In these models the structure of the disk 
(in both vertical and radial directions)  
is calculated in detail.
The radiation from the rotating and infalling envelope
 is calculated 
from detailed envelope models (CHKW, HCB), in which the transfer 
equation is solved at each frequency including both scattering
and emission by dust. The envelope calculations from CHKW and HCB assume 
that the luminosity comes from a star and a disk; in the case 
of HL Tau, $\sim$ 95\% of the 
luminosity is estimated to originate from the disk.

DCH have shown that irradiation of a circumstellar disk by an optically-thick
infalling envelope that reprocesses the disk radiation 
can quantitatively explain the high outer disk temperatures
required by the submillimeter and millimeter observations of HL Tau. The 
envelope heating, 
determined
by modeling completely independent constraints 
(the Spectral Energy Distribution (SED)  between
$1 \mu$m $\lesssim \lambda \lesssim 100 \mu$m; CHKW, HCB),  
dominates the outer disk temperature distribution.

The disk is assumed steady, geometrically thin, in vertical
hydrostatic equilibrium and 
with a turbulent viscosity calculated 
using the standard $\alpha$ viscosity prescription  (Shakura \&
 Sunyaev 1973\markcite{sha73}), 
i.e., the viscosity coefficient is expressed as
$\nu = \alpha  c_{\rm s}  H$,
where $c_{\rm s}$ is the local sound
speed, $H$ is the local pressure scale height of the gas and $\alpha$
is the viscosity  parameter  (assumed constant
throughout the disk). 
For an $\alpha$ disk, 
the surface density of mass $\Sigma$ is given by

\begin{equation}
\Sigma={\Mdot \Omega_K \over 3 \pi \alpha c_s^2(T_c)}
\end{equation}
where $\Mdot$ is the disk mass accretion rate, $\Omega_K$ is the 
Keplerian angular velocity, and $T_c$ is the midplane temperature. 

Since $\Sigma \propto T_c^{-1}$, the higher the midplane temperature 
(which is controlled by the envelope irradiation)
 the smaller the surface density. For the mm continuum, the opacity is 
dominated by dust and is independent on temperature.
Thus, the continuum optical depth  is $\tau_\nu \propto T_c^{-1}$, and 
 the higher the envelope irradiation flux, 
the smaller the $\tau_\nu$.
The source function at mm wavelengths is proportional to $T_c$. This means 
that as long as the envelope heats the $\alpha$ disk enough 
 to make the outer regions ($R > 50 \ \AU$) optically thin in the 
mm continuum, 
 the brightness temperature distribution  in this spectral range 
is independent on $T_c$.
  From a practical point of view, this implies that the mm continuum 
brightness distribution and emergent flux of the disk are not very 
sensitive to the details of the adopted envelope model.
 However, a modified sheet-collapse envelope 
was assumed because it has a reduced envelope extinction
along polar directions as required to account for the observed scattered
light nebulae
(HCB).

DCH find that disk models characterized by 
\begin{equation}
\biggl ( {\Mdot \over 10^{-6} \ \MSUNYR} \biggr ) 
\biggl ( {\alpha \over 0.04} \biggr )^{-1} 
\biggl ( {R_d \over 125 \AU} \biggr )^{1/2} \approx 1
\end{equation}
(where $R_d$ is the outer radius of the disk)
 have the same submillimeter to radio SED than HL Tau.
The interferometric visibility at $\lambda=0.87$ $\mu$m (from 
Lay \et 1994\markcite{lay94}) constrains the disk radius to be 
$110 \ \AU \lesssim R_d \lesssim 140 \ \AU$ and the inclination angle 
between the disk axis and the line of sight, $i=60^\circ$.
Considering that the bolometric luminosity of HL Tau, $L_{BOL} \approx 5 \ \LSUN$, 
 is 
dominated by accretion (Calvet \et 1994), i.e., $L_{BOL} \approx 
L_{acc} = G \Mdot M_* /R_*$, 
 and that the central star is a  typical T Tauri object 
with $M_*=0.5 \ \MSUN$ and $R_*=3 \ \RSUN$, we adopt a disk 
 accretion rate $\Mdot=10^{-6}$ $\MSUNYR$. From the family of models 
which fits the observed properties of HL Tau, we have selected the 
one  with a viscosity parameter 
$\alpha=0.04$, a disk outer radius $R_{\rm d}=125$ AU.
This is a physical model constructed to explain observed properties of 
HL Tau, which might not be those of a typical embedded disk. 
Using the  correlation between accretion luminosity and the 
luminosity of the $Br\gamma$ line in CTTS, Muzerolle, 
Hartmann, \& Calvet (1998\markcite{muz98}) infer that Class 
I sources have similar accretion luminosities than Class 
II sources. This result implies mass accretion rates for Class I 
sources of $10^{-8}$-$10^{-7} \ \MSUNYR$ (with a median value 
$10^{-7} \ \MSUNYR$ according to 
Calvet, Hartmann, \& Strom 1999\markcite{cal99}) , and 
that the disk of HL Tau is  denser by one or two orders of magnitude 
than typical disks.
However, we chose this model to present our molecular line calculations 
because in the case of HL Tau  there are enough continuum 
observations to constrain the disk and the envelope physical properties. 
In \S 5.3 we discuss the effect of changes in the disk
temperature and density, on the properties of the molecular lines. 


A preliminary study of molecular line emission from 
protoplanetary disks
was presented in G\'{o}mez \& D'Alessio\markcite{gom95} (1995). 
In that paper we
assumed a disk of negligible thickness, and therefore no vertical
structure was considered. Only the radial power-law dependence of 
temperature and
density that fit better to observational data was used. With those
assumptions, we did not consider some opacity effects that 
have an important influence on line profiles and intensities (see
\S\ref{results}). In the improved
calculations we are presenting now, these effects caused by the disk
vertical structure are taken into account properly.

\section{Integration of the Transfer Equation for Line Emission}

\subsection{Simplifying assumptions}
\label{assumptions}

Besides the assumptions described in \S\ref{structure} to derive
the disk structure, we assume LTE for the population of the molecular
energy levels. 
The particle density is $n > 10^6$ cm$^{-3}$ 
for regions
of the disk with heights over the disk midplane $z\lesssim 4
H$, and reaches values of $n\simeq 10^{16}$ cm$^{-3}$ in some parts of the
midplane. 
Therefore, the regions of the
disk that can contribute significantly to the molecular line emission
studied here have densities that are higher than the critical
density of thermalization for all the molecules considered in this
paper, making the LTE assumption acceptable.

We also consider thermal line profiles. The other major contributor to
line profiles could be turbulence. However, it is difficult to
estimate a particular turbulent velocity, since for a given $\alpha$
parameter, the corresponding velocity varies greatly depending on the
origin we assume for the turbulence, e.g., hydrodynamic
(Shakura, Sunyaev, \& Zilitinkevich 1978\markcite{sha78};
Zhan 1991\markcite{zha91}; Dubrulle 1992\markcite{dub92}),
or magnetohydrodynamic (Balbus \& Hawley 1991\markcite{bal91}; Hawley,
Gammie, \& Balbus 1995\markcite{haw95}). 
Given this indetermination, we first included just the thermal
contribution in the line profiles, and studied afterwards (\S
\ref{turbulence}) how the
computed spectra change if we include a turbulent velocity equal to
the sound speed. 
It is unlikely that the typical velocity of the turbulent eddies is
supersonic since in this case the turbulent motions would be
dissipated by shocks (Frank, King, \& Raine 1992\markcite{fra92}).

\subsection{The System of Equations}
\label{equations}

Once we know the detailed density and temperature distribution
throughout the disk, to determine the emission from any given molecular
transition we just have to integrate the corresponding transfer equation
\begin{equation}
\frac{dI_\nu}{ds} = \kappa_\nu (S_\nu - I_\nu) \; , \label{transfereq}
\end{equation}
where $I_\nu$ is the intensity, $\kappa_\nu$ is the absorption
coefficient, and $S_\nu$ is the source function. We note that for a
proper calculation, $I_\nu$ should include line emission from the gas,
plus continuum emission from the dust of the disk. The way to do this
is to consider $\kappa_\nu = \kappa_l + \kappa_c$, with contributions
from both line and continuum.

In practice, the integration of Eq.\ \ref{transfereq} along the line
of sight is performed by solving numerically the equations
\begin{equation}
I_\nu = I_{\rm bg} e^{-\tau_{\rm t}} + \int_0^{\tau_{\rm t}} S_\nu
  e^{-\tau} d\tau
\end{equation}
\begin{equation}
\tau = \int_{s}^{\infty} (\kappa_l + \kappa_c) \rho ds \; ,
\end{equation}
where $I_{\rm bg}$ is the background intensity, 
$s$ is the distance along the line of sight 
(the observer is located at $s = +\infty$, and $s=0$ corresponds to the
disk midplane; see Appendix \ref{coordinates}  for a description of 
the coordinate systems
used here), $\tau$ is the optical depth between the observer and point
$s$, $\tau_{\rm t}$ is the
total optical depth, and $\rho$ is the mass density of the gas. 
We can reduce this system of integral equations
to the equivalent differential ones
\begin{equation}
\frac{dI'}{ds} = -S_\nu (\kappa_l+\kappa_c) \rho e^{-\tau} \label{eqint}
\end{equation}
\begin{equation}
\frac{d\tau}{ds} = - (\kappa_l + \kappa_c) \rho \label{eqtau}
\end{equation}
with $I' = I_\nu - I_{\rm bg} e^{-\tau_{\rm t}}$. We solved this
system of equations numerically using a Runge-Kutta algorithm with
variable step size (Press \et 1992\markcite{pre92}).

\subsection{Absorption Coefficients}
\label{opacity}

The particular model we assumed for the disk structure will be
included  in the density, in the
absorption coefficients $\kappa_l$ and $\kappa_c$, which depend on
temperature, and in the source function that, since we are
assuming LTE, can be expressed as the 
Planck function,  $S_\nu=B_\nu(T)$. 

The absorption coefficient for the continuum 
is dominated by dust, and can be written as:
\begin{equation}
\kappa_c = 0.07 \left(\frac{\nu}{1.5\times 10^{12}}\right)\; ,
\label{kappac} 
\end{equation}
with a 
dependence on frequency  consistent 
with the slope of the SED of HL Tau, for $\lambda \gtrsim  500$
\micron\ (e.g., Beckwith \et 1990\markcite{bec90}; Beckwith \& Sargent
1991\markcite{bec91}; DCH).
The coefficient in Eq. \ref{kappac} is obtained assuming that at frequencies 
lower than $\nu = 1.5 \times 10^{12}$, the dust opacity is 
given by Draine \& Lee (1984\markcite{dra84}).

On the other hand,
the absorption coefficient for 
a molecular transition in the gas is 
\begin{equation}
\kappa_l = \frac{A_{\rm ij} c^2}{8\pi\nu^2} \frac{g_{\rm i}}{g_{\rm j}}
  \left[ 1- \exp\left(-\frac{h\nu}{kT}\right) \right] \frac{n_{\rm j}}{\rho}
  \Psi(v) \; ,\label{kappal}
\end{equation}
where $A_{\rm ij}$ is the Einstein coefficient for the $i\rightarrow
j$ transition, $g_{\rm i}$ and $g_{\rm j}$ are the statistical weights of
the upper and lower states respectively, $n_{\rm j}$ is the density of
molecules in the lower state, and $\Psi(v)$ is the line profile, which
we assume as thermal. These last two parameters take the form
\begin{equation}
\frac{n_{\rm j}}{\rho} = \frac{X_{\rm mol}}{m_\circ} g_{\rm j}
\exp\left(-\frac{E_j}{kT}\right) Q^{-1} \label{density} 
\end{equation}
\begin{equation}
\Psi(v) = \frac{c}{\nu} \left(\frac{m_{\rm mol}}{2\pi k
    T}\right)^\frac{1}{2} \exp\left[-\frac{m_{\rm mol}
    (v-v_\circ)^2}{2kT} \right] \; ,\label{profile}
\end{equation}
with $m_\circ$ the mean molecular mass of the gas 
(that we assume to be 2.5
times the proton mass),  
$m_{\rm mol}$ the mass of the molecule, $X_{\rm mol}$ its molecular
abundance relative to
hydrogen, $Q$ is the partition function, and
$v_\circ$ is the macroscopic velocity of the gas, which for the
purpose of this integration is the component of the Keplerian
velocity on the line of sight 
(see Appendix \ref{projected}). Molecular parameters are given in
Table \ref{moleculartab}.


Substituting Eqs. \ref{kappac}, \ref{kappal},
\ref{density}, and \ref{profile} in Eqs.\ \ref{eqint} and \ref{eqtau},
we can now integrate 
these last two equations for the  
density
and temperature structure of a disk model, 
along any line of sight.

\subsection{Gridding and Convolution}
\label{conv}

We performed the integration of the transfer equation for every line
of sight defined by the fine grid of cells
that is detailed in Appendix \ref{grid}, 
 and for successive line-of-sight
velocities. In this way we obtained the intensity spectrum in each
cell. We used a convenient coordinate system for the integration,
which  is described in Appendix \ref{coordinates}.
 
With this intensity distribution, we can then convolve with an arbitrary
beam to obtain the flux observed with the corresponding angular
resolution as
\begin{equation}
F_\nu(X,Y,v) = \sum_{\rm beam} I_\nu(x,y,v) A(x,y) P(x,y)
\label{flux}
\end{equation}
i.e., the sum over the beam of the intensity ($I_\nu$), times the cell
area ($A$), times the Gaussian beam pattern ($P$), assuming a Gaussian
with peak value of unity at position ($X,Y$). Therefore, the flux
$F_\nu$ is the observed spectrum in any given sky position $X$, $Y$,
in units of Jy beam$^{-1}$.

\section{Results}
\label{results}

\subsection{Line Detectability}
\label{ldetect}

We have calculated the line intensity expected for the model of disk
structure in HL Tau defined in \S\ref{structure}. This has been
done for several transitions of different molecules. The results are
shown in Table \ref{sensit_tab04}, where it can be seen the maximum
flux for each molecular line when convolved to simulate observations
with an angular resolution
of 0\farcs4, which would allow to obtain resolved images of disks. We
chose this particular beam size as a trade-off between resolution and
sensitivity  because it is enough to resolve a
disk of $\sim 0\farcs 7$ (100 AU) radius, while it is large enough to
give a good signal-to-noise ratio. 
We also compare the model results with the sensitivity
of several interferometers (both presently working and in project)
that can achieve this angular resolution, after 10 hours of
integration time.


In view of this table, we can draw several conclusions in terms of
detectability of this type of disks with subarcsecond resolution. First, the
presently working interferometers (e.g., VLA for \nh\ lines, VLA for
CS with 13 receivers at 7 mm) seem unable to detect resolved
protoplanetary disks using a reasonable 
amount of observing time. This
is consistent with the non-detection of \nh\ lines 
(G\'{o}mez \et 1993\markcite{gom93}).
 On the other hand, future interferometers (e.g., SMA, MMA)
could reach such a detection. The SMA sensitivity seems just enough to
detect these lines, while the MMA could reach a good signal to noise
in a few minutes of observing time.  It is obvious that for this
particular type of observations, large collecting areas in the 
interferometers are critical to guarantee a real chance of success.
An instrument like the MMA would
be an important tool for these studies, since it would obtain detailed
maps of several species and transitions in reasonable observing times,
thus allowing to test and constrain models of disks and their
structure. Note that in the MMA case, the S/N ratio is so good that it
will be possible to detect lines form 
protoplanetary disks with a finer angular
resolution. 

In Table \ref{sensit_tab3} we show the expected flux of line emission
for an angular resolution of $3''$, compared with the sensitivity of
some presently working interferometers, which routinely reach that
resolution. As mentioned above, observations with an
angular resolution on this order cannot provide the resolved image of a
disk, but some indirect studies can still be made in the cases where
the telescope sensitivity is enough for a detection of line emission.


Another important issue when we try to map molecular lines from disks
is to choose the appropriate molecular species and transitions for
these studies. The best lines will be those at whose frequency the material
around the disk-star system (a dense envelope or a clump in which the star
may be embedded) is optically thin. This is mainly to avoid absorption
of the disk emission by the enveloping material, which could be
critical for sources in early stages of evolution, as in the case we
are studying here. Possible confusion
induced by emission from the enveloping material is less important,
since interferometers tend to resolve out large-scale structures. With
the mentioned condition, it would be best to use low abundance
molecular species, and high excitation molecular transitions, that
trace selectively the disk gas with respect to the cooler cloud
material surrounding the star-disk system. 
However, when trying for high excitation transitions, we
cannot go too high in frequency, since absorption due to dust in the
envelope will start to be important. Some kind of compromise should be
reached here, and it will depend on the column density of the envelope.

Among the molecular transitions shown in Table \ref{sensit_tab04},
those of C$^{17}$O could be good candidates, since they are probably
optically thin for the ambient gas. It is interesting (and fortunate) 
that the expected
fluxes for the C$^{17}$O lines are higher than the corresponding ones
of C$^{18}$O, even though the former is $\sim 4$ times less abundant than
the latter. 
This is so because the transitions of the less abundant
species becomes optically thick deeper into the disk, and the inner
parts of the disks are hotter than the outer ones. 

More common CO isotopes are likely to be reliable probes in optically
visible T Tauri stars, although it is significant to find that a rare
isotope as C$^{17}$O can also be detected, which is specially
important for embedded objects.

\subsection{Spectra and maps of line emission}
\label{lmap}

To illustrate in detail the kind of results that we can obtain in 
subarcsecond observations of disks, we chose to  concentrate in the
following on C$^{17}$O lines, since they  appear as well-suited
candidates.
Spectra and maps of other molecules are qualitatively similar to the C$^{17}$O
ones. Quantitative differences are illustrated by the values given 
in Tables \ref{sensit_tab04}  and \ref{sensit_tab3}. 

Detailed maps and spectra obtained from our calculations in the
particular case of the  C$^{17}$O($3\rightarrow2$) and 
($2\rightarrow1$) transitions are
shown in Figures \ref{c17o3_chmap}, \ref{c17o2_chmap},
\ref{c17o3_integ}, \ref{c17o2_integ},  \ref{c17o3_spec}, 
\ref{c17o2_spec},  
\ref{c17o3_posvelra}, and \ref{c17o3_posveldec}. These results 
also correspond to the HL Tau structure
model described
in \S\ref{structure}, and they are convolved to a 
resolution of 0\farcs 4.

The channel maps in Figs.\ \ref{c17o3_chmap} and \ref{c17o2_chmap} 
are plotted with the
projected major axis of the disk lying along a line of constant
declination. With the assumed inclination angle of 60\arcdeg, the
southern half of the disk is closer to the observer than the northern
one. The central star is located at position (0,0).


In these maps, we can see several characteristics that eventually can
be tested observationally. First, we see a clear north-south
asymmetry; the areas of the disks that are farther away from the
observer show a more intense line emission than those that are
closer. This is an expected behavior for optically thick lines in a
disk of finite thickness. In
these cases, for the same projected distance from the star, the lines
of sight intersect the disk surface closer to the star (thus tracing
warmer gas) in those areas that are inclined farther away from the
observer. This effect can also be seen in the integrated line intensity
(Fig.\ \ref{c17o3_integ}). If this asymmetry is actually observed, it could
be compared with observations of jets and outflows, to see if they are
consistent with each other. For instance, redshifted lobes of outflows
should be projected in the same direction as the part of the disk closest
to us, assuming that disks and outflows are perpendicular. 
The east-west asymmetry that can be seen at
$v=0$ \kms (Fig.\ \ref{c17o3_chmap}) is due to the asymmetric 
hyperfine structure of
the C$^{17}$O lines.


We can also see in Fig.\ \ref{c17o3_chmap}, that the peak emission for
the lowest velocities tend to trace the outer edge of the disk.  
One of the reasons for the
molecular line flux to trace the outer parts of the disk (as already
pointed out by Sargent \& Beckwith 1991\markcite{sar91}) 
is that the effective area
emitting at a definite velocity within the telescope beam 
increases with distance from the star, and
more steeply than an eventual decrease of brightness temperature with
distance. This  effect 
is also the reason of the  double peaked profiles that is
typical of line emission from rotating structures. 

This geometrical 
effect is further reinforced by the opacity of the dust
continuum emission. 
If the continuum optical depth tends to infinity, the brightness
temperature for both line+continuum and continuum alone will be equal
to the kinetic temperature at the disk surface. 
Therefore, 
there would be no contrast between the line and the continuum emission, and the
line would be undetectable. Thus, the effect of high continuum
opacities is to effectively reduce the brightness temperature of the line
(after subtracting the continuum), i.e., to reduce the contrast between
line and continuum. This decrease of brightness temperature
is important as we
approach the parts of the disk closer to the star (with higher densities,
and therefore with higher continuum opacities), but not in the outer parts,
where
the continuum is usually optically thin. Moreover, in a real disk with high
(but not infinity) continuum optical depth close to the star,
the opacity is higher at
frequencies with line emission (due to the contribution of line+continuum)
than at those with continuum emission alone. Therefore, the continuum
emission gets optically thick deeper into the disk than the line+continuum
emission. As the temperature increases as we move vertically towards the
disk midplane, the molecular lines could be seen {\em in absorption} in these
high opacity regions (see Figs. \ref{c17o3_spec}, \ref{c17o2_spec}).


One of the results of this opacity effect is that the integrated intensity
map (Figs. \ref{c17o3_integ} and \ref{c17o2_integ}) 
shows the appearance of a ring, with lower emission at the
center. When detecting such a ring-like structure, one could be tempted to
wrongly conclude that the lower emission corresponds to a lower 
amount of gas, as
in a toroidal gas distribution. This result illustrates that it is important
to properly consider the contribution of the continuum emission from 
dust in these calculations of line
emission, a contribution that sometimes is ignored.

\subsection{Position-Velocity Diagrams}
\label{posvel}

Figs. \ref{c17o3_posvelra} and \ref{c17o3_posveldec} show
position-velocity diagrams along lines with constant
declination and right ascension, respectively. Superimposed to those
diagrams there are lines that mark the corresponding mid-plane 
Keplerian velocity at
each position, taking into account the distance from the star and the
inclination angle of the disk. We see that the emission peaks in the cut
along the major axis of the projected 
disk (Fig. \ref{c17o3_posvelra}, top left) would be enough to 
give us an
estimate of the mass of the central star, provided that we have
independently determined
the inclination angle of the disk. In fact, molecular line
observations can help in this determination, 
although it is not
straightforward (see
\S\ref{angle}).
The fit of the other position-velocity diagrams to the Keplerian lines is
not as good as for the cut along the major axis, due to the finite angular
resolution ($0\farcs4$).
We could try a more refined fit of the position-velocity diagram to
the lines of
Keplerian velocities, by looking whether the emission maxima at each
velocity are relatively close to the Keplerian lines 
(see, e.g., Sargent \& Beckwith 1991\markcite{sar91},
for the large-scale structure in HL Tau). However, as we can see by taking a
close look at Fig. \ref{c17o3_posvelra} (top left), those emission 
maxima are close to the
Keplerian expectation at high relative velocities (with absolute value
higher than 2 \kms\ in this particular case), 
but depart significantly from it at the lowest velocities, for which the
emission maxima are located closer to the star than in the corresponding
Keplerian line.


\section{Effects of changes in physical parameters}
\subsection{Inclination angle}
\label{angle}

An obvious question to ask is what qualitative and quantitative effect
would be observable if a disk with the physical structure of that of
HL Tau is located to an inclination angle with respect to us different
from the $60^\circ$ assumed here, and obtained from the continuum
observations in HL Tau. To illustrate these effects, we show in Figs.
 \ref{c17o3_30d_chmap} and \ref{c17o3_30d_spec}  
the results of our calculations with an inclination angle
of $30^\circ$, of the C$^{17}$O($3\rightarrow2$) transition.


In quantitative terms, the maximum intensity for $i=30^\circ$ is 325
mJy beam$^{-1}$ (with a beam of 0\farcs4), which is 25\% higher than
the maximum intensity for the same transition in the disk at 
$i=60^\circ$. 
Qualitatively, an important difference can be seen in the spectra (Fig
\ref{c17o3_30d_spec}, to be compared with Fig \ref{c17o3_spec}). The 
double peak in the central
spectrum is much less prominent for $i=30\arcdeg$, while for
$i=60\arcdeg$ there was a significant self-absorption. Therefore, 
the combination
of lower radial velocities and lower optical depth in less inclined
disks makes this central spectral feature very sensitive to the value
of the inclination angle. 
It would be difficult to use this absorption feature alone to
estimate the
inclination angle of the disk, because a similar self-absorption can
be produced by strong vertical temperature gradients. 
However, it can be an important piece of information that, together
with other observational evidences, can be used to self-consistently
obtain the disk structure and the inclination angle, in a similar way
as explained in sec. \ref{structure}. 

The observed aspect ratio of the integrated intensity (Fig.\ 
\ref{c17o3_integ}) 
is somewhat correlated with the inclination angle, but it does not directly
indicate its value. If we measure the distance between integrated intensity
 peaks on each
side of the central star, for both the major and minor axes, to obtain
this aspect ratio, it would give an apparent inclination angle of 
$\sim 30^\circ$ for $i=60^\circ$, and $\sim 20^\circ$ for $i=30^\circ$. 
Therefore, the derivation of the inclination angle from geometrical
considerations must be made with care.

We also note that the maxima of emission at each velocity (see
Fig. \ref{c17o3_30d_chmap}) does not follow the outer radius of 
the disk, as it did for
the disk at $i=60\arcdeg$

\subsection{Turbulent velocity}
\label{turbulence}

In this section, we show how the above calculations change 
when  there is a turbulent velocity  field, 
 in addition to thermal and rotational motions of the gas.
To check for the maximum possible effect that the inclusion of
turbulence will have on the line spectra, we use an upper limit to
the turbulence velocity, which we assume equal to the thermal velocity
(see \S\ref{assumptions}).

The turbulent velocity modifies the profile given by
Eq. (\ref{profile}). 
If we assume a Gaussian distribution for the turbulent velocity field,
the resulting profile will also be Gaussian, and can be written as

\begin{equation}
\Psi(v) = \frac{c}{\nu} \left( {1 \over \zeta_\circ \pi^{1/2}} \right)
     \exp\left[-\frac{
    (v-v_\circ)^2}{\zeta_\circ^2} \right] \; ,\label{turb_prof}
\end{equation}
where
\begin{equation}
\zeta_\circ ^2 =  {2kT \over m_{\rm mol} } + \zeta_T^2
\end{equation}
and $\zeta_T$ is the 
turbulent velocity dispersion
we have assumed equal to the 
thermal velocity dispersion 
of a particle with 
a mass equal to the mean molecular mass $m_o$,    
 i.e., $\zeta_T^2 = {2kT \over m_{o}}$ (see \S\ref{assumptions}).

A comparison between Figs. \ref{c17o3_chmap} and
\ref{c17o3_turb_chmap} 
shows that the
line intensity increases (roughly by a factor 3, for the
turbulent 
velocity adopted in our calculations) 
  if  a turbulent velocity is included 
 in the line profile calculation. 
 The main reason of this enhancement of the emission 
is that the line opacity decreases when  
 the velocity of the emitting molecules increases (see
Eqs. (\ref{kappal}) 
and (\ref{turb_prof})). 
Thus,  the ``turbulent lines'' are dominated by emission 
from a deeper, hotter region than those involved 
in the intensity of a ``pure thermal line''.

From an observational point of view, this enhancement of the intensity
of the lines when turbulence is considered, will obviously make the
detection of these lines easier. Thus, the conclusions one can draw
from Table 2 in terms of detectability should be considered as the
less favorable case, since that line intensities in a non-turbulent
scenario are lower limits to the lines intensities with some turbulent
contribution. 


The mean increase in linewidth is only of $\sim 0.4$
\kms\ when applying a turbulent component at sound speed. This means
that the line profile is dominated by rotational motion, which makes it
difficult to determine the turbulent velocity from line
profiles. Such a determination will require a high spectral
resolution and high sensitivity, 
but it may be necessary if we want to derive physical
characteristics of disks from line emission. Since turbulence causes
strongest lines, it could be confused by the effect that a different
density and/or temperature structure may cause.

\subsection{Generalization to other sources}

In this paper we have used a disk structure that is consistent with
observational data for the relatively well constrained disk of HL
Tau. An important issue is whether our conclusion that such a disk
will be detectable with millimeter interferometers is of general
applicability to other sources, with different disk structures in
density and temperature. This is not an easy question to address,
since to derive a realistic disk structure will require a wealth of
data that, at present, 
is probably only available for HL Tau (see
sec. \ref{structure}). However, we can obtain some answers by checking
how changes in 
temperature and density affect a possible detectability of molecular
line emission. 
The range of values we have explored goes from $T_\circ/5$ to
$5T_\circ$ and from $n_\circ/100$ to $100 n_\circ$, where $T_\circ$
and $n_\circ$ are the temperature and volume density of the original HL Tau
model (Fig.\ \ref{model}).
The resulting maps are
very similar morphologically to the ones obtained above for the HL Tau
model. Spectral signatures and diagnosis mentioned in the
previous sections (e.g., asymmetry of the line emission, enhancement
of line emission towards the edges, tracing of Keplerian velocities,
changes of central self-absorption as a function of inclination angle,
emission enhancement with turbulent motions)
are of general applicability in these parameter
ranges, which include typical Class I sources (see below).

The main result is that the peak intensity of the maps is much more
sensitive to temperature than to density variations. 
For instance, for the C$^{17}${O(J$=3\rightarrow 2$) lines,
the relation of intensity with temperature is linear, while the volume density
dependence is as shown in Fig.\ \ref{densdep}, with variations of less
than a factor of 5 over 4 orders of magnitudes in density.
The conclusion in terms of detectability is
that one should tend to observe the hottest disk sources to ensure an
easier detection.


We can further focus on the detectability of disks in Class I sources
in general.
In the context of the model of an accretion disk irradiated by 
an infalling envelope, the density scales as $\Mdot/\alpha$ (c.f. \S 2) 
and the temperature 
is controlled by the emergent flux from the envelope. The last one 
 depends on the input luminosity provided by the central star and 
the inner disk (i.e., $L_*$ and $L_{acc} \propto \Mdot$).
 Taking the envelope irradiation flux 
proportional 
to the input luminosity (Calvet, private communication), the   irradiation 
temperature is $T_{irr} \propto L^{1/4}$, which is equal to the 
disk photospheric temperature for $R \gtrsim 1 \ \AU$ (DCH). 
In the case of HL Tau, the input luminosity is $L=5 \ \LSUN$.
 However, the median luminosity of Class I sources is $\sim 1 \
\LSUN$ 
(e.g., Kenyon \& Hartmann 1995). Thus, the disk of a typical Class I source 
would have a photospheric temperature 0.7 times 
the temperature we 
calculate for the HL Tau model. 
On the other hand, Muzerolle, Hartmann \& Calvet  
(1998\markcite{muz98}) estimate 
that typical  
Class I sources 
have mass accretion rates similar to classical T Tauri stars, 
$\Mdot \sim 10^{-8} - 10^{-7} \ \MSUNYR$, with a median $\Mdot=10^{-7} \ \MSUNYR$ (Calvet, Hartmann \& Strom 1999\markcite{cal99}). The density of such a disk
is a factor of $100-10$  
lower than the density of the HL Tau disk model, assuming the 
same viscosity parameter $\alpha$.  
Combining the dependence of flux on 
both density (as shown in Fig.\ \ref{densdep}) and
temperature (linear), and using $n = n_\circ/100$ and 
$T=0.7T_\circ$,
we estimate that the molecular lines of a typical Class I source 
would have at worst 0.18 times the intensity we have calculated 
for the HL Tau model. Thus, the worst case implies that 
 the C$^{17}${O(J$=3\rightarrow 2$) lines of Class I 
sources would be detectable in a reasonable observing time with the 
MMA, but are probably not detectable using SMA.

\subsection{Molecular abundances}

An issue of concern is the possibility of molecular depletion
within the disk, since one would think that lower molecular abundances will
make molecular lines more difficult to detect. 
In the calculations presented in this paper, we have used a constant
abundance, which in most cases is the interstellar one (for CS we
assumed an abundance given in Blake et al. (1992\markcite{bla92}) 
for the disk in HL Tau, 
lower than the interstellar one).
Theoretical models of the evolution of molecular abundances in 
protoplanetary disks predict
depletion of CO from the gas phase for temperatures $ T  \ <$ 20 K 
(Aikawa \et 1996 \markcite{aik96}) and of NH$_3$, for $T \ < $ 80 K (Aikawa et
al. 1997\markcite{aik97}).  The application of these results to 
the HL Tau model is not straightforward, since there are differences between 
the  physical conditions of their disk model and ours. 
However, since the most relevant disk property in determining 
molecular abundances is the temperature (Aikawa \et 1996\markcite{aik96}), 
and for $R > 10$ AU our model has a radial distribution of photospheric 
temperature similar to the model used by Aikawa \et
(1996\markcite{aik96}, 1997\markcite{aik97}) 
we adopt some of their results in this discussion.
The HL Tau disk model (with $R_d=125$ AU) has a temperature higher 
than the critical temperature 
for freezing $\sim 20 $ K (reached at $R \simeq 200$ AU, 
in the disk model of Aikawa \et 1996\markcite{aik96}, 
1997\markcite{aik97}). On the other hand, the timescale for transforming 
CO into CO$_2$ ice 
 at $ T \simeq 30 $ K is larger than $ \sim 1 $ Myr 
(Aikawa \et 1996\markcite{aik96}), and the viscous timescale of our 
disk model is $\sim 0.1 $ Myr (DCH).  
 Therefore, depletion is probably not
significant for CO lines (which are our best candidates for detection) 
in the conditions of our HL Tau model. 

For typical Class I sources, with temperatures 0.7 times that of HL
Tau, we have checked that the maximum intensity of
C$^{17}$O lines is the same with and without considering depletion, 
for different beam sizes (0\farcs 4,
0\farcs 6, 0\farcs 8). Therefore, depletion does not seem to be
important regarding
line detectability in typical Class I sources.
For even colder sources, CO
depletion could, in principle, affect detectability if a significant
fraction of the disk is below the critical temperature of 20 K.

Depletion of NH$_3$ from the gas phase  at $T < 80$ K ($R \gtrsim$ 20 AU) 
will probably 
decrease the intensity of the ammonia lines with respect to our 
calculations, confirming our 
conclusion that these 
lines are not good candidates for detecting protoplanetary disks.  

Blake \et (1992\markcite{bla92}) find that CS is depleted in the 
disk of HL Tau respect 
to the molecular cloud, and we have used the 
lowest abundance they have estimated  
(see Table 1) for calculating the CS and C$^{34}$S lines. 
However, Blake \et (1992\markcite{bla92}) observations do not resolve the inner 
$\sim$ 100 AU region of HL Tau, and 
we could be underestimating the abundance of CS. 
Thus, we calculate the maximum intensity of the line CS(J$=1\rightarrow 0$) 
assuming an abundance $X_{mol} =5 \times 10^{-9}$ (Aikawa \et 1996\markcite{aik96}), 
i.e., a factor of 25 larger than the abundance quoted in Table 1.
The maximum intensity is $F_\nu = 9.6$ mJy beam$^{-1}$, 
for a beam size of $0\farcs4$, which is still too low to be detected in 
a reasonable observing time given the sensitivity of the VLA.

 In any case, the study of molecular
abundances in disks will certainly benefit from accurate observational
data, which can be obtained with the type of observations we are
trying to reproduce. The results of our model can also help to
interpret observational results in terms of abundance changes.

\section{Conclusions}

In this work we model the expected line emission from 
a protoplanetary disk irradiated by an infalling envelope, 
when observed with subarcsecond resolution. 
We adopt a model for the  
disk structure 
consistent with the available observational constraints for 
HL Tau. Our main conclusion is that 
the  detection of molecular lines at 
subarcsecond resolution 
will be within reach of the projected millimeter and 
submillimeter interferometers
(e.g., MMA, SMA).
 In particular, an instrument like the MMA has enough 
sensitivity to provide 
 important data to constrain disks models.

We suggest that the lines C$^{17}$O($2 \rightarrow
1$) and ($3 \rightarrow 2$),  at $\lambda=$1.335 
and 0.89 mm, respectively,    
are good candidates for detecting disk lines at subarcsecond
resolution, 
because the infalling envelope  is probably optically thin at the 
corresponding wavelengths  (CHKW).
Also, these lines are more intense 
than others from more abundant species (e.g., C$^{18}$O) 
due to opacity effects combined  with the 
temperature gradients in the disk.

There is a clear asymmetry in the line intensity, with more 
intense emission in the disk area farther away from the observer. This
can be directly used to compare the geometrical relationship between
disks and outflows. A decrease of intensity towards the
center of the disk is also evident.

The emission peaks in position-velocity diagrams lie on the lines that
trace mid-plane Keplerian velocities. This can be used to determine
the stellar mass, but only if and independent estimate of the
inclination angle of the disk can be obtained.

Some changes in physical properties are correlated with changes in
line intensity and spectral shape:
\begin{itemize} 
\item Line intensity
decrease as the inclination angle increases. 
\item For larger
inclination angles, a central absorption feature in the spectra becomes
deeper.
\item Increasing the turbulent velocity results in brighter lines, but only
in a moderate enhancement of linewidth.
\item The line intensity scales linearly with disk temperature, but is
less sensitive to density changes.
\end{itemize}
Going the opposite way, i.e., obtaining physical properties of disks
from line shape and intensity is not straightforward, since a similar
observational characteristic can be explained by several alternative
physical changes. However, the information provided by molecular line
observations can eventually 
be used, together with continuum data, to obtain a
self-consistent model that could explain all observational evidence.

Expected changes in molecular abundances do not affect our results
regarding the detectability of molecular lines from disks, at least
for typical Class I sources.

We plan to extend this study to models of disks in a later state of
evolution, not surrounded 
by an infalling envelope but irradiated  by the central star. A 
detailed calculation 
of how the line properties depend on disks parameters could be a useful 
disk diagnostic tool.

\acknowledgments

We thank J. M. Torrelles, L. F. Rodr\'{\i}guez, J. Cant\'{o}, 
S. Lizano, and our referee, J. Najita for their useful comments.
We also thank J. Ballesteros for his help preparing the figures.
We acknowledge the support of CONACyT (Mexico) and CSIC (Spain), 
which funded part of this work. 
JFG is supported in part by DGICYT grant PB95-0066, by INTA grant 
IGE 4900506, 
and by Junta de Andaluc\'{\i}a (Spain). JFG also thanks 
the Instituto de Astronom\'{\i}a 
of the Universidad Nacional Aut\'{o}noma de M\'{e}xico for their
hospitality during the
preparation of this paper. PD thanks the Insituto de Astrof\'{\i}sica 
de Andaluc\'{\i}a for their hospitality, a DGAPA scholarship 
and support from CONACyT project J27748E 
during part of this work.

\appendix

\section{Some details on the integration process}

\subsection{Projected Velocities}
\label{projected}

Fig.\ \ref{figprojected} shows the geometry of the problem, and how
to calculate Doppler shifts of the gas. A gas element at radius $r$
from the central star rotates with a Keplerian velocity
\begin{equation}
v=\sqrt{\frac{G M_*}{r}}\; .
\end{equation}
At a particular point on its orbit, defined by angle $\theta$, the
projected velocity along the line of sight, for an inclination angle
$i$, is
\begin{equation}
v_\circ=\sqrt{\frac{G M_*}{r}}\cos \theta \sin i \; .
\end{equation}


\subsection{Grid}
\label{grid}
To integrate the transfer equation, we divided the disk into a grid of
cells. The transfer equation is then solved for a line of sight going
through the center of each cell. To create this grid, we first
calculate the lines on the disk midplane with the same projected
velocity, which are of the form
\begin{equation}
r=\frac{G M_*}{v^{2}_{\circ}} \cos^2 \theta \sin^2 i \; .
\end{equation}
These lines are shown in Fig.\ \ref{figgrid}a. 

Up to a radius of 50 AU, the grid is defined so that the borders of
the cells are 
these isovelocity lines and their perpendicular
lines (of the form $r=A\sqrt{(\sin\theta)}$, where $A$ is a constant),
so that all
cells have the same velocity resolution, that we choose to be lower
that the thermal velocity width of the gas. This grid avoids a possible
overestimate of the fluxes in the inner disk at the velocities we
sample. This overestimate would happen if
cells include a range of projected velocities larger than the
thermal velocity of the gas. An example of the grid we chose is shown in Fig.\
\ref{figgrid}b. For radii larger than 50 AU, the cells on this
grid have a large area. This makes the sampling of temperature and
density poorer on those areas and therefore we chose a
Cartesian grid, with smaller cells than those corresponding to the
grid shown, thus keeping the velocity range covered by each cell below
the thermal velocity width of the gas.


\subsection{Coordinate Systems}
\label{coordinates}

In this section, we define the coordinate system used in the
integration of the transfer equation. We found this coordinate system
to be convenient for an easy transformation of all the
equations. Fig.\ \ref{figcoordinates} illustrates the coordinate
systems mentioned in this section.


We first consider the ``disk'' coordinate system $(x, y, z)$, centered on
$O$ (the position of the central star). The axes $x$ and $y$ are on
the disk plane. Axis $z$ represents the height from the disk
midplane. Equations for temperature and density of the disk are given
naturally on this coordinate system. 

We can also consider the ``sky'' coordinate system $(x_s, y_s, s)$,
centered on $O$. Axis $x_s$ coincides with $x$. Axes $x_s$ and $y_s$
are on the plane of the sky. Axis $s$ is along the line of sight.

The system  we use for the integration (``integration'' coordinate
system) is $(x_s, y_s,
s')$, centered on $O'$. It is just a translation of the ``sky''
system along axis $s$, so that coordinates on axis $s'$ equal zero
when the line of sight intersects the disk midplane. Note that there
is actually a different coordinate system for each line of sight,
i.e., for every point on the sky where we integrate the transfer
equation, we use a different coordinate system.

The transformation between the ``disk'' and ``integration'' coordinate
systems is then

\begin{eqnarray}
y= \frac{y_s}{\cos i}-s'\sin i \\
z=s' \cos i \;.
        \label{eqtransform}
\end{eqnarray}

Equations that define the physical characteristics of the disk are
then transformed into the ``integration'' coordinate
system. Integration of the transfer equation is performed in two
parts, from
$s'=+\infty$ to 0, and from 0 to $-\infty$, to properly sample the
disk midplane.

\clearpage

\begin{figure}
\plotone{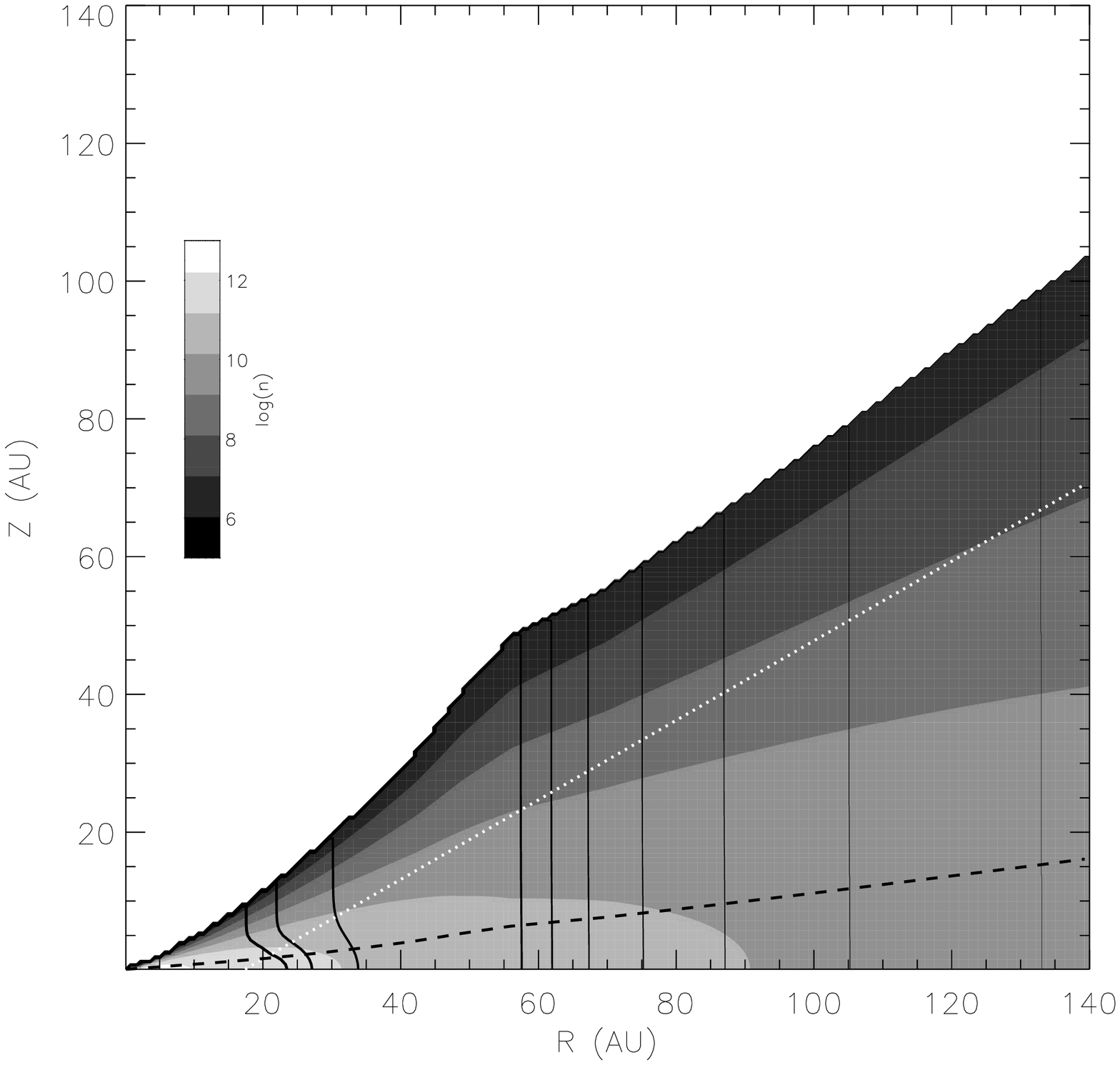}
\caption{\label{model} Map of temperature (Solid contour
lines) and  number density of 
hydrogen molecules (greyscale) 
for a model of an accretion disk irradiated by an infalling envelope. The disk 
has $\Mdot=10^{-6} \ \MSUNYR$ and $\alpha=0.04$, and surrounds a 
star with $M_*=0.5 \ \MSUN$ and $R_*=3 \ \RSUN$. The envelope is 
flat, with $\Mdot_{env}=4 \times 10^{-6} \ \MSUNYR$ (see DCH and
HCB for details). 
The increment step for temperature contours is 5 K and
the lowest temperature
plotted (the rightmost one) is 25 K, with thicker contours indicating 
larger temperature values.
The gas scale height is shown as a heavy dark dashed line, and the
line of sight, assuming an inclination angle $i=60 \ ^\circ$ for the
disk axis,
is plotted with a dotted white line.}
\end{figure}

\begin{figure}
\epsscale{0.8}
\plotone{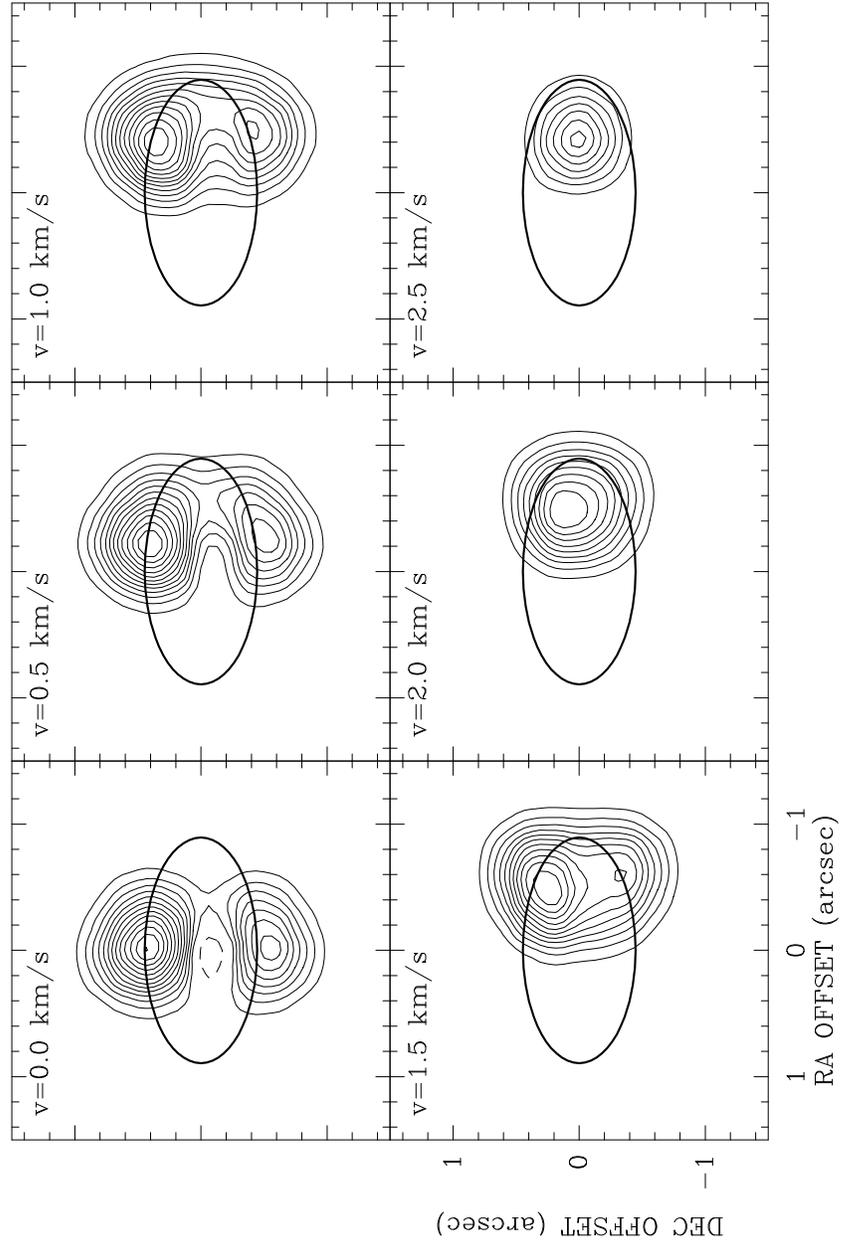}
\caption{\label{c17o3_chmap} Channel maps of
C$^{17}$O($3\rightarrow2$) line emission, for a disk with
$i=60^\circ$, and thermal line profiles, convolved with a beam of
0\farcs4. The lowest positive contour and the increment step are 20
mJy beam$^{-1}$. 
The velocity with respect to
the rest system of the central star is shown at the top left corner of
each panel. The thick ellipse corresponds to the disk maximum radius 
projected on the plane of the sky.}
\end{figure}

\begin{figure}
\epsscale{0.8}
\plotone{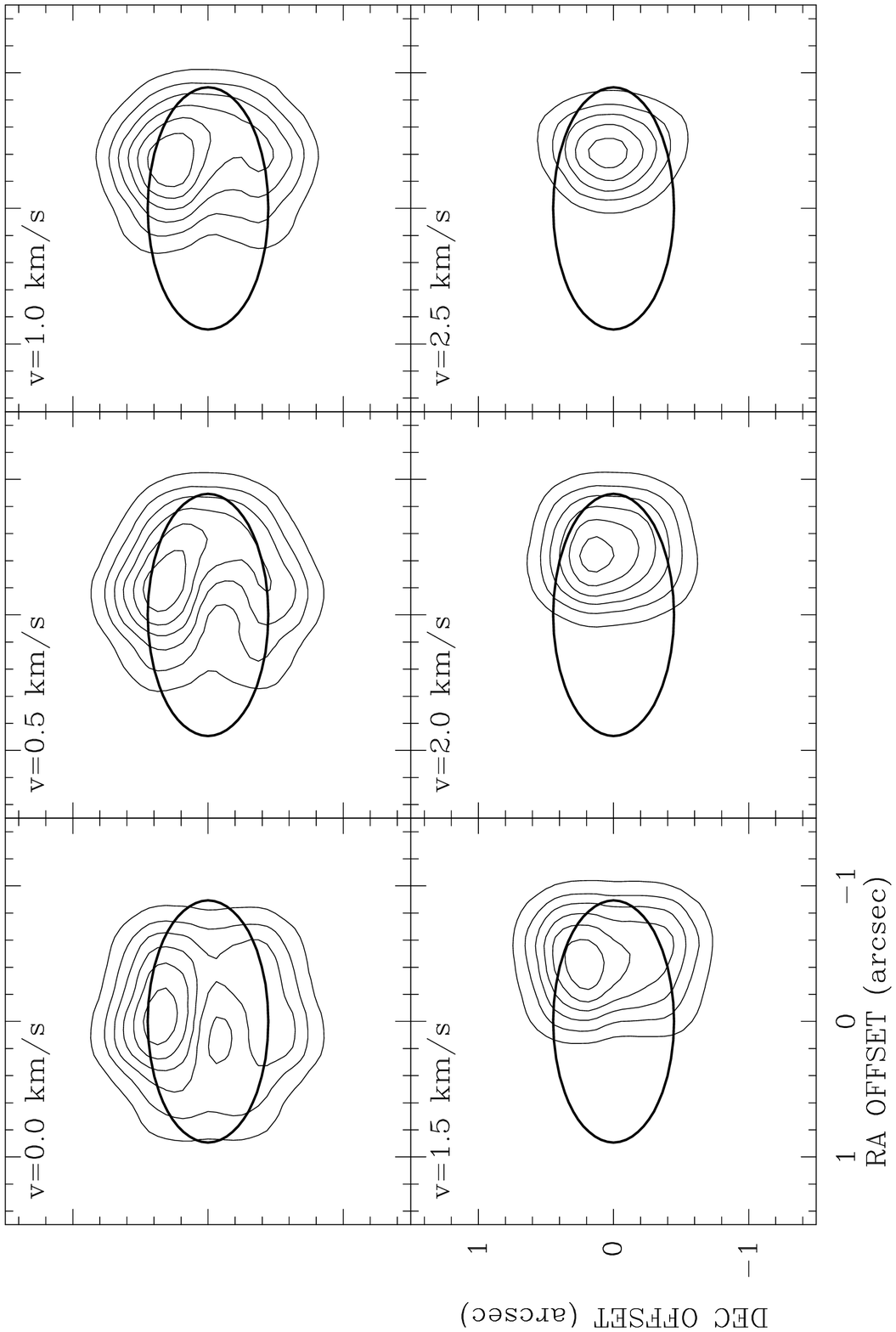}
\caption{\label{c17o2_chmap} Same as Fig.\ \ref{c17o3_chmap}, but
for C$^{17}$O($2\rightarrow1$) line emission.}
\end{figure}

\begin{figure}
\plotone{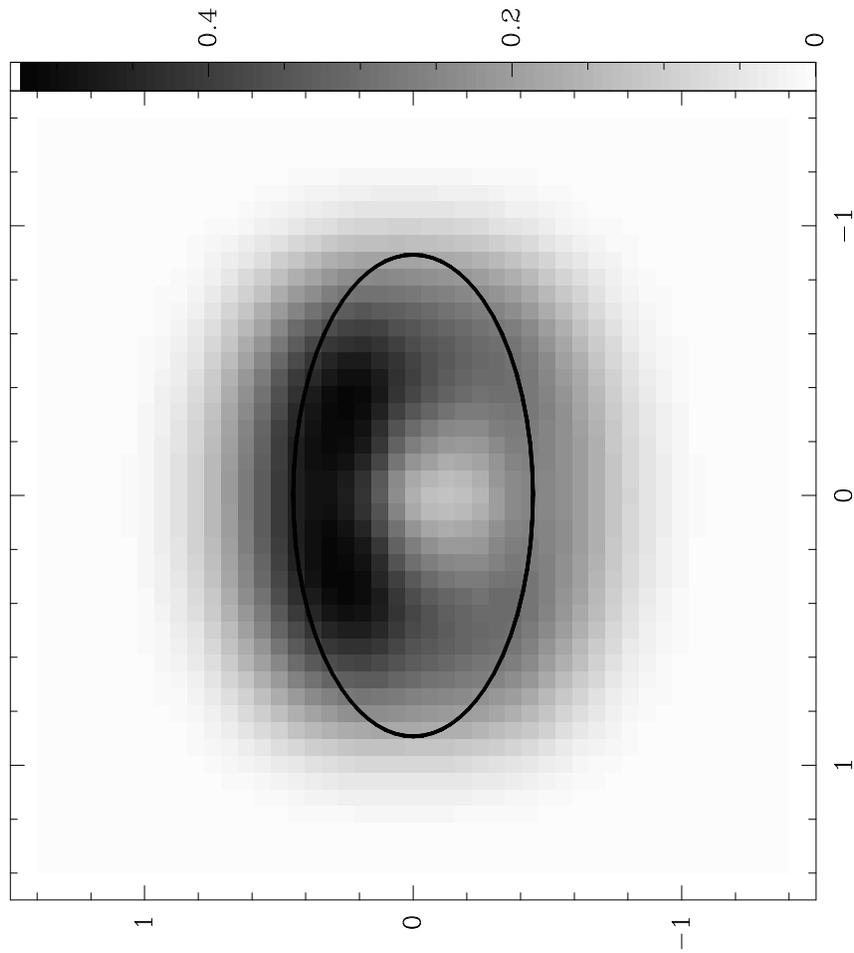}
\caption{\label{c17o3_integ} Gray scale map of the integrated
intensity of C$^{17}$O($3\rightarrow2$) for a disk with
$i=60^\circ$, and thermal line profiles, convolved with a beam of 
0\farcs4.}
\end{figure}

\begin{figure}
\plotone{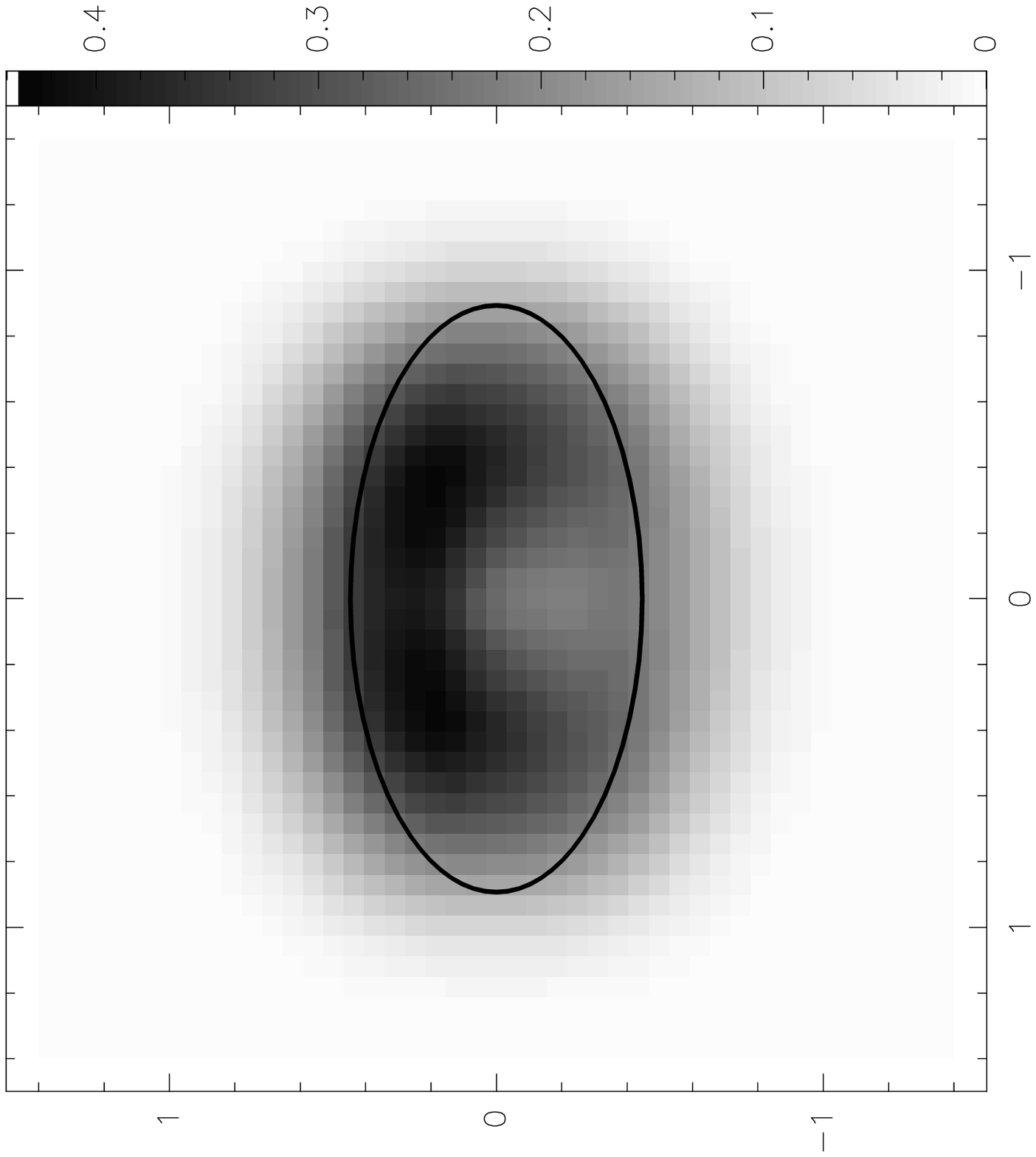}
\caption{\label{c17o2_integ} Same as Fig.\ \ref{c17o3_integ}, but
for C$^{17}$O($2\rightarrow1$) line emission.}
\end{figure}

\begin{figure}
\plotone{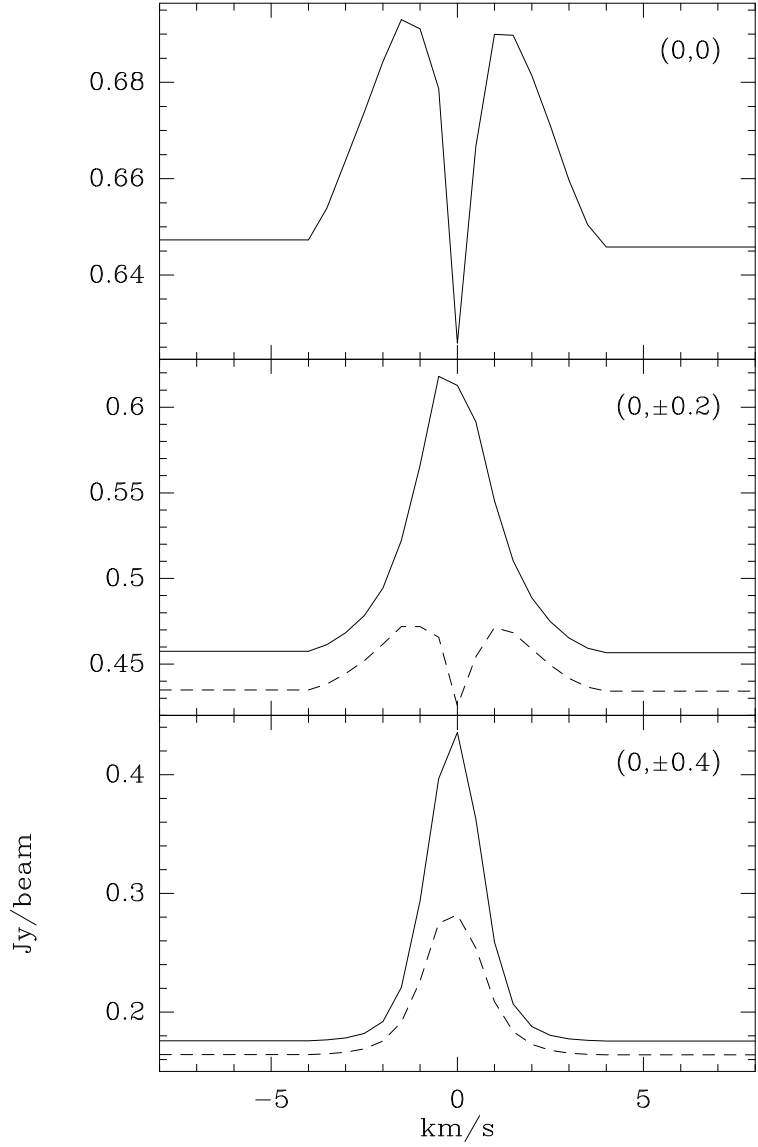}
\caption{ \label{c17o3_spec} Spectra of the 
C$^{17}$O($3\rightarrow2$) line emission at selected positions, for a disk with
$i=60^\circ$, and thermal line profiles, convolved with a beam of 
0\farcs 4. Position
offsets, in arcseconds, with respect to the central star 
are indicated at the top right corner of each panel. Solid and dashed
lines correspond to the spectra at positive and negative declination
offsets, respectively.}
\end{figure}

\begin{figure}
\plotone{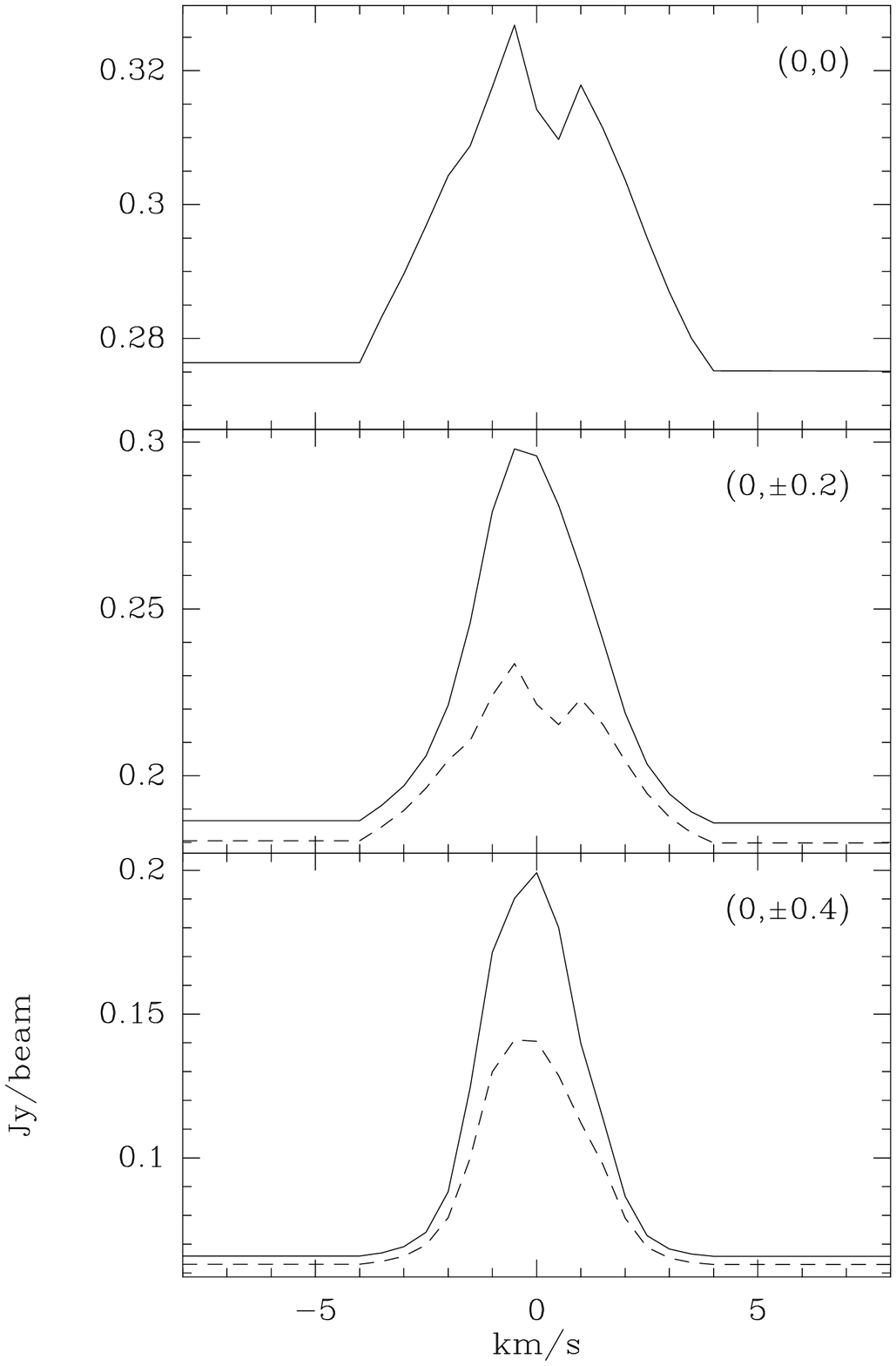}
\caption{\label{c17o2_spec} Same as Fig.\ \ref{c17o3_spec}, but
for C$^{17}$O($2\rightarrow1$) line emission.}
\end{figure}

\begin{figure}
\plotone{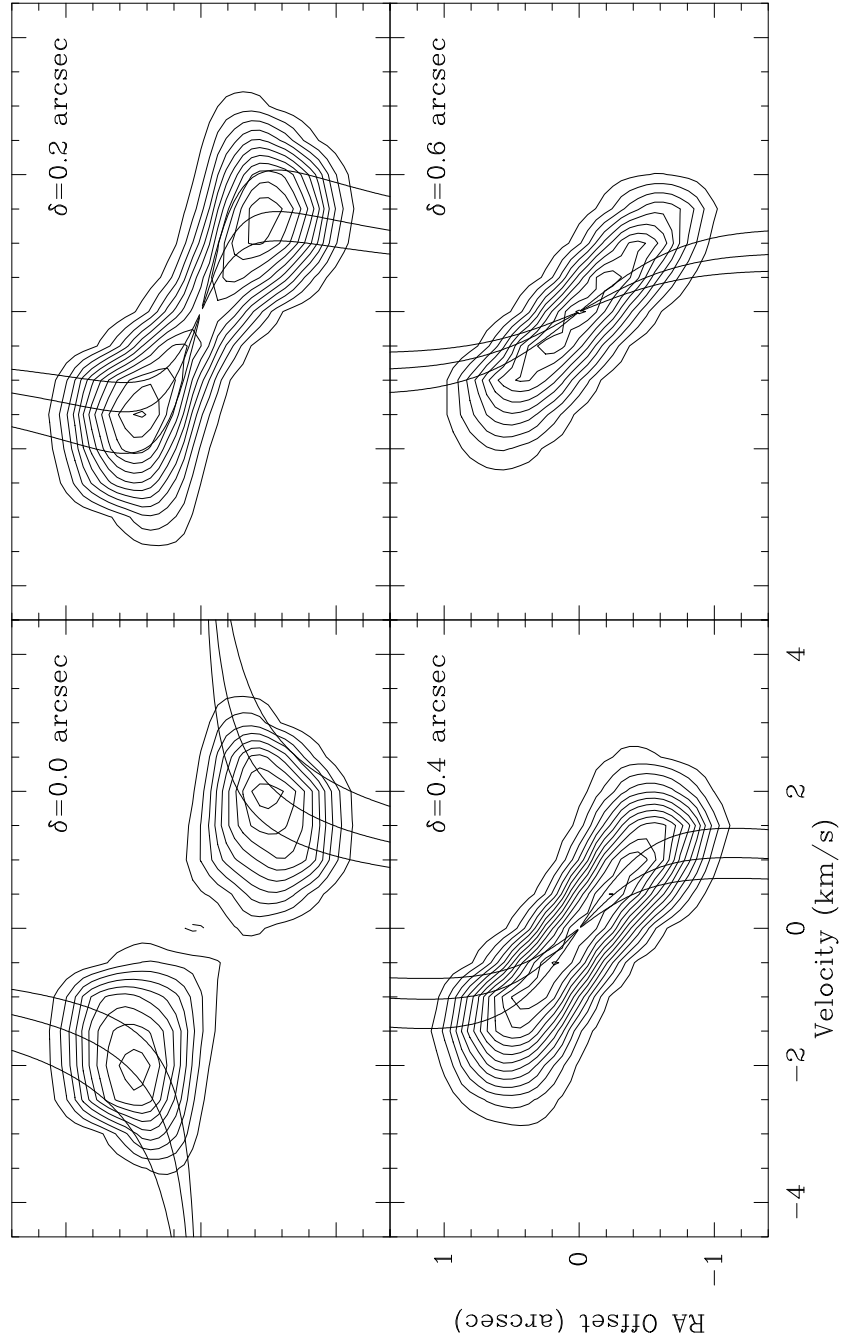}
\caption{\label{c17o3_posvelra} Position-velocity diagrams of 
C$^{17}$O($3\rightarrow2$) emission, along lines of constant
declination (indicated at the top right corner of each panel), i.e.,
parallel to the mayor axis of the projected disk, assuming
$i=60^\circ$ and thermal line profiles, and convolved with a beam of 
0\farcs 4. The lowest positive contour and the increment step are 20
mJy. Thin solid lines
correspond to Keplerian laws, for different values for the mass of the
central star, $M_*=0.25, \ 0.5,$ and $1 \ \MSUN$. The disk model 
constructed for 
HL Tau, assumes  that the central star has $M_*=0.5 \ \MSUN$.}
\end{figure}

\begin{figure}
\plotone{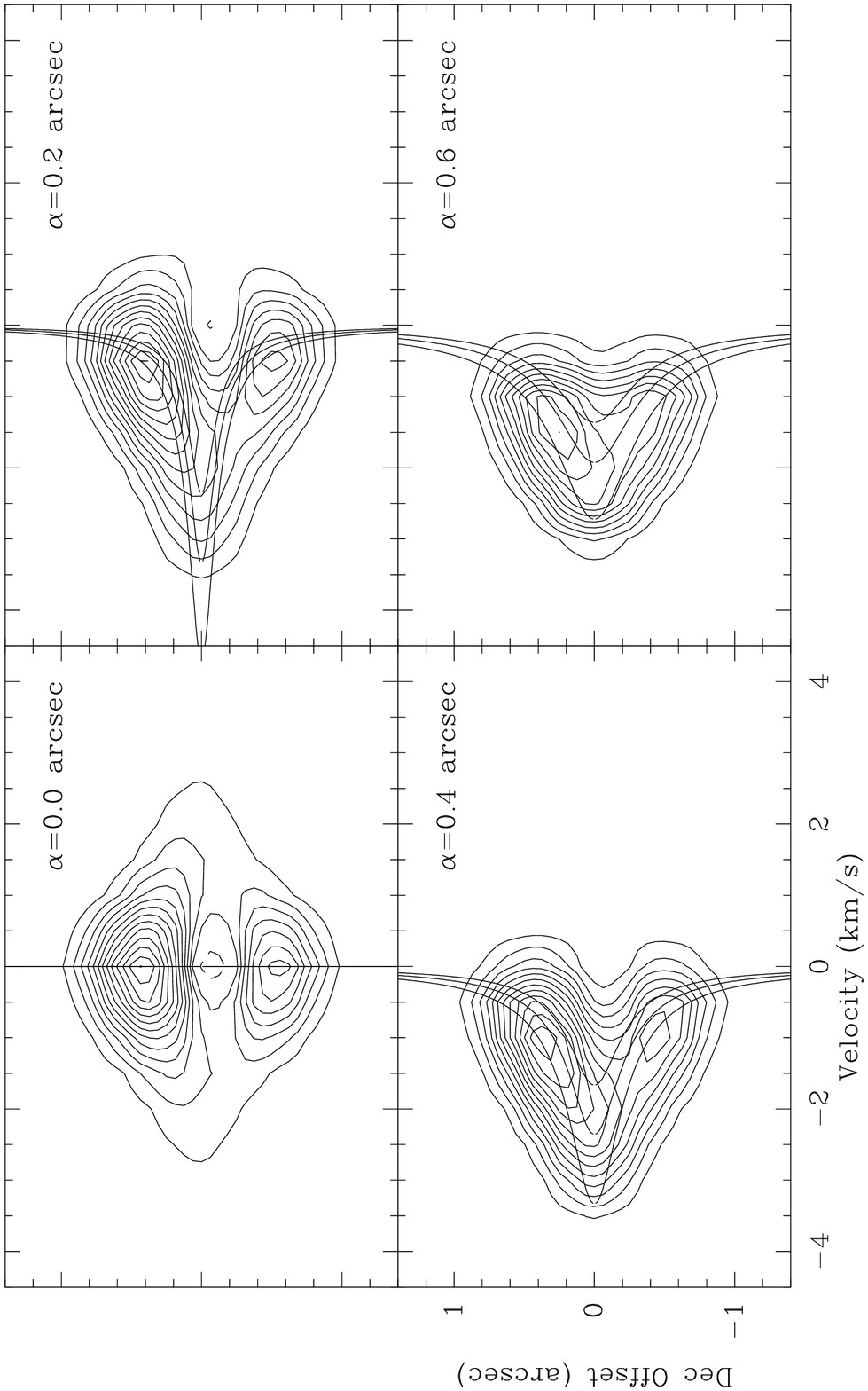} 
\caption{\label{c17o3_posveldec} Position-velocity diagrams of 
C$^{17}$O($3\rightarrow2$) emission, along lines of constant
right ascension (indicated at the top right corner of each panel), i.e.,
parallel to the minor axis of the projected disk, assuming
$i=60^\circ$ and thermal line profiles, and convolved with a beam of 
0\farcs 4. The lowest positive contour and the increment step are 20
mJy beam$^{-1}$. Thin solid lines
correspond to Keplerian laws at the disk midplane, 
for different values for the mass of the
central star, $M_*=0.25, \ 0.5,$ and $1 \ \MSUN$. The disk model 
constructed for 
HL Tau, assumes  that the central star has $M_*=0.5 \ \MSUN$.}
\end{figure}

\begin{figure}
\plotone{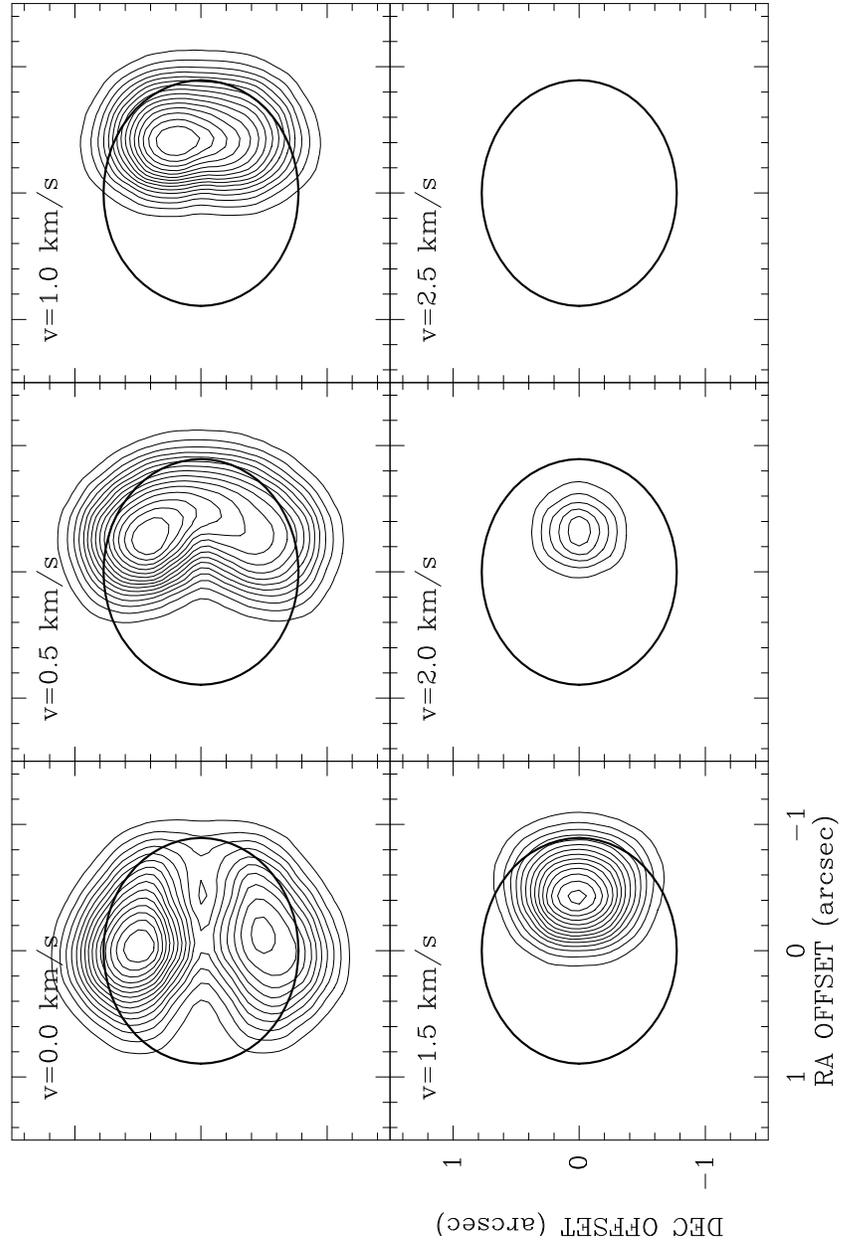}
\caption{\label{c17o3_30d_chmap} Same as Fig.\ \ref{c17o3_chmap},
but for an inclination angle $i=30^\circ$.}
\end{figure}

\begin{figure}
\plotone{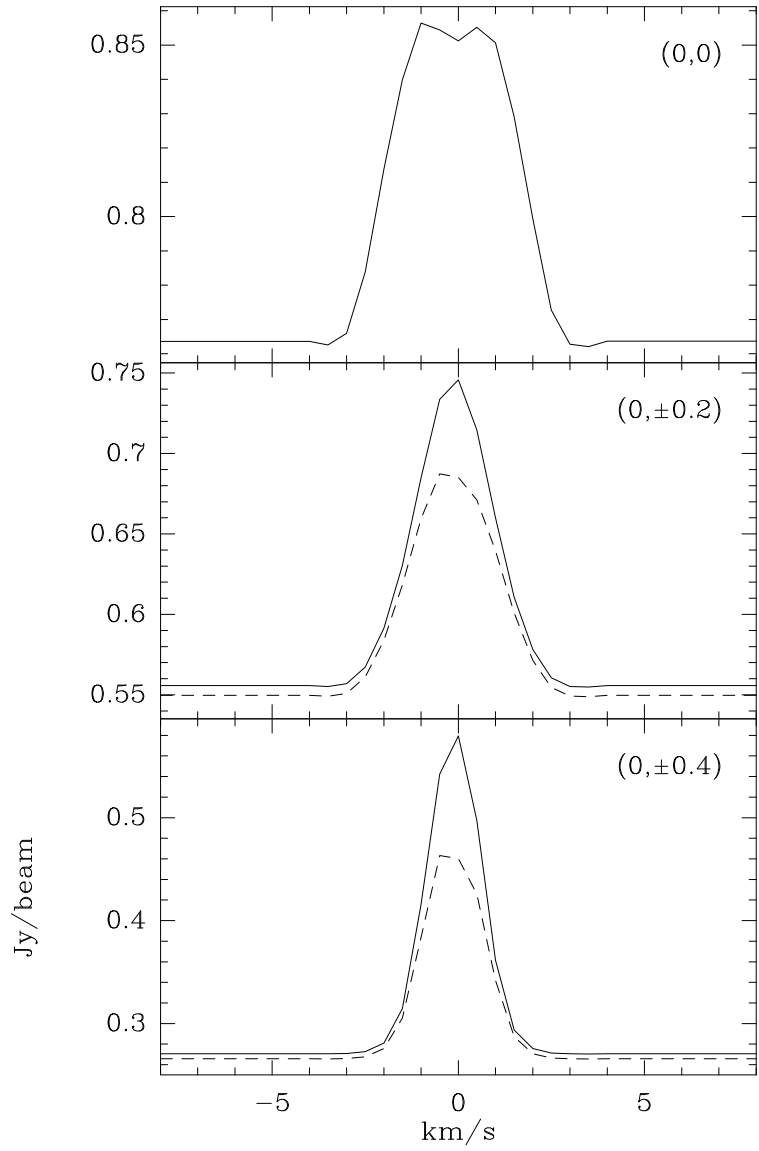}
\caption{ \label{c17o3_30d_spec} Same as Fig.\ \ref{c17o3_spec}, 
but for $i=30^\circ$.}
\end{figure}

\begin{figure}
\plotone{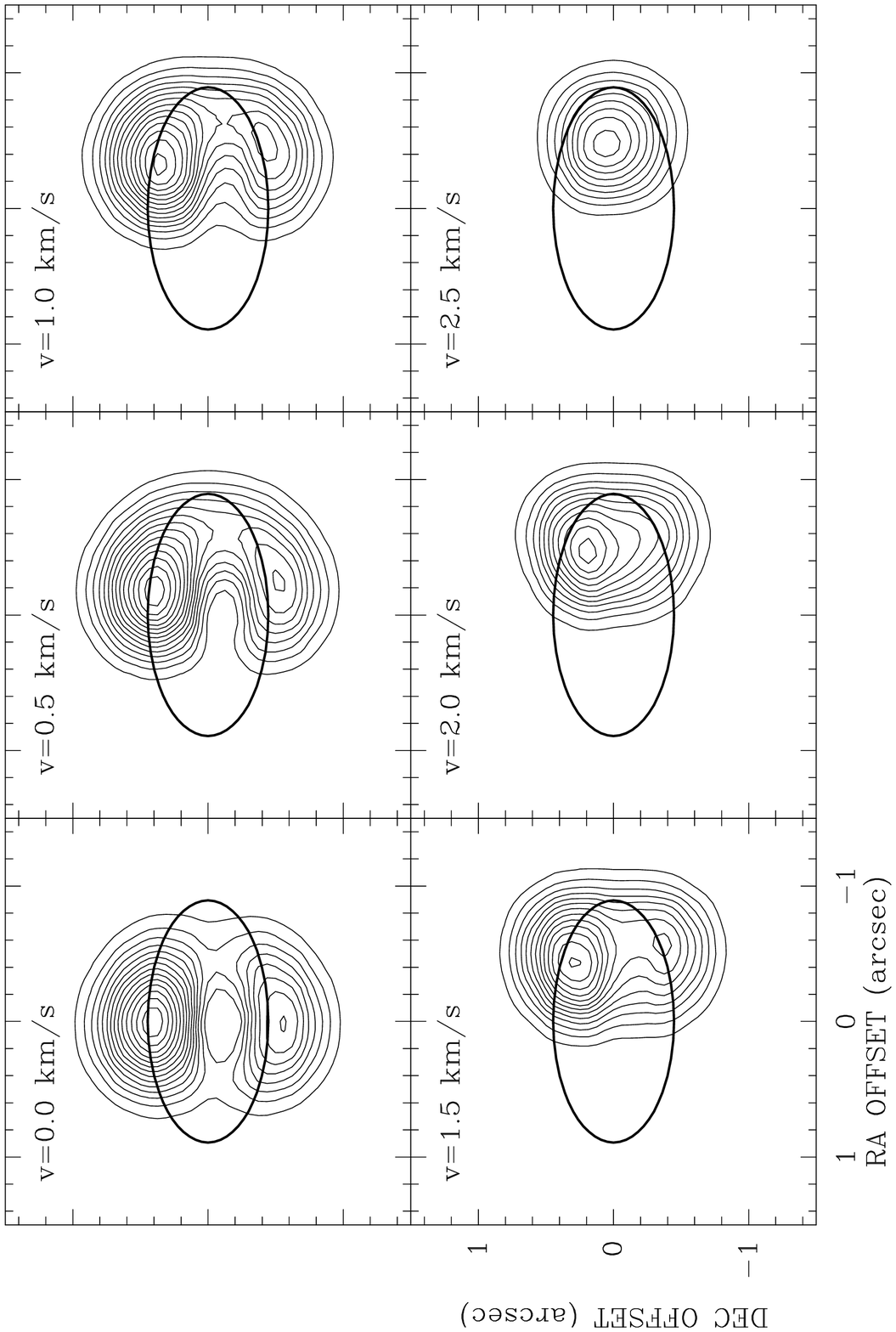}
\caption{ \label{c17o3_turb_chmap} Channel maps of
C$^{17}$O($3\rightarrow2$) line emission, for a disk with
$i=60^\circ$, including a turbulent component for the line profile and
convolved with a beam of
0\farcs 4. The lowest positive contour and the increment step are 20
mJy beam$^{-1}$. The velocity with respect to
the rest system of the central star is shown at the top left corner of
each panel. The thick ellipse corresponds to the disk maximum radius 
projected on the plane of the sky.}
\end{figure}

\begin{figure}
\plotone{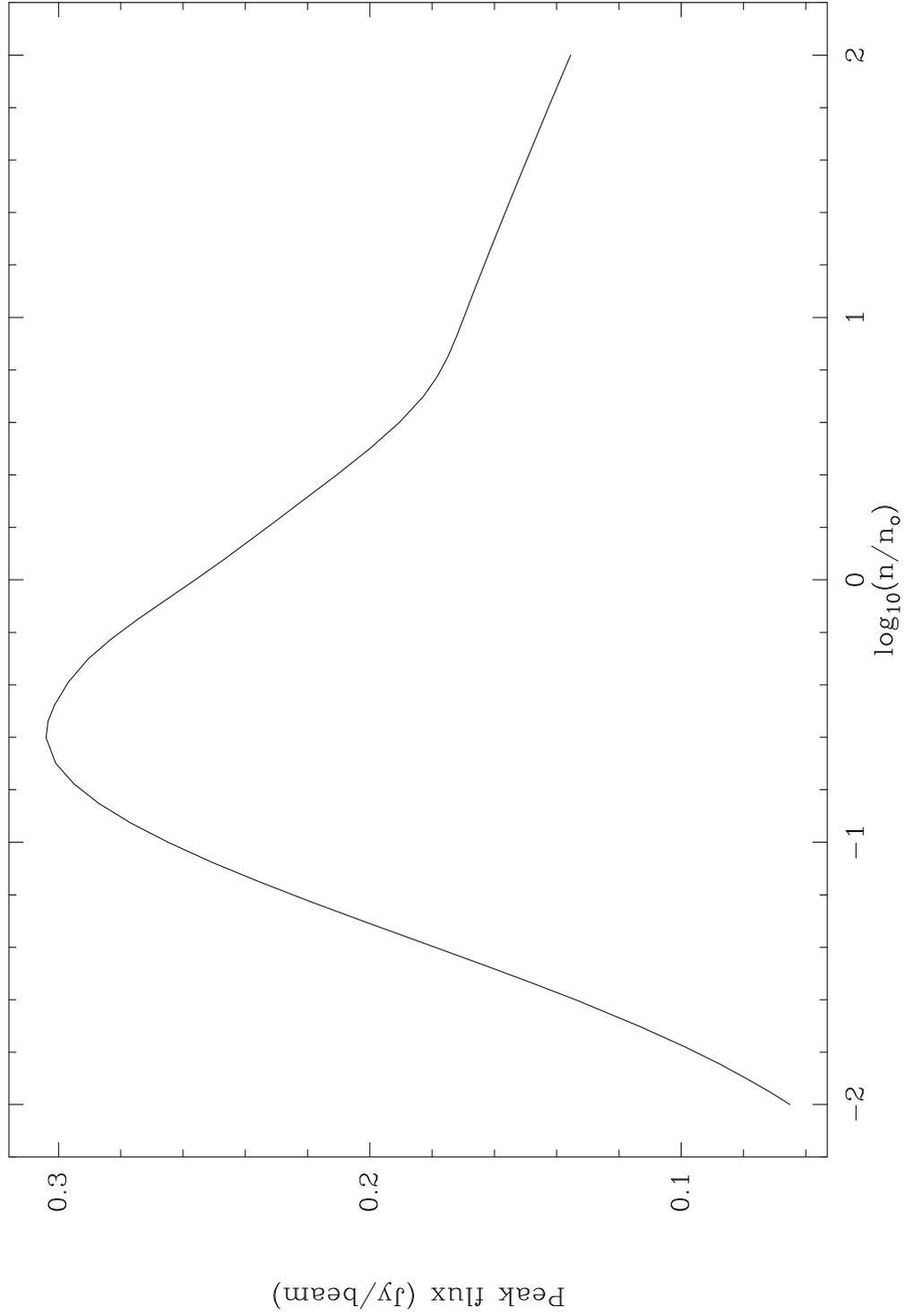}
\caption{\label{densdep} Peak flux of the
C$^{17}$O($3\rightarrow2$) line as a function of disk density, for
$v=1$ \kms. The x-axis is the logarithm of the ratio between the
density and the original density model ($n_\circ$) used for HL Tau.}
\end{figure}

\begin{figure}
\plotone{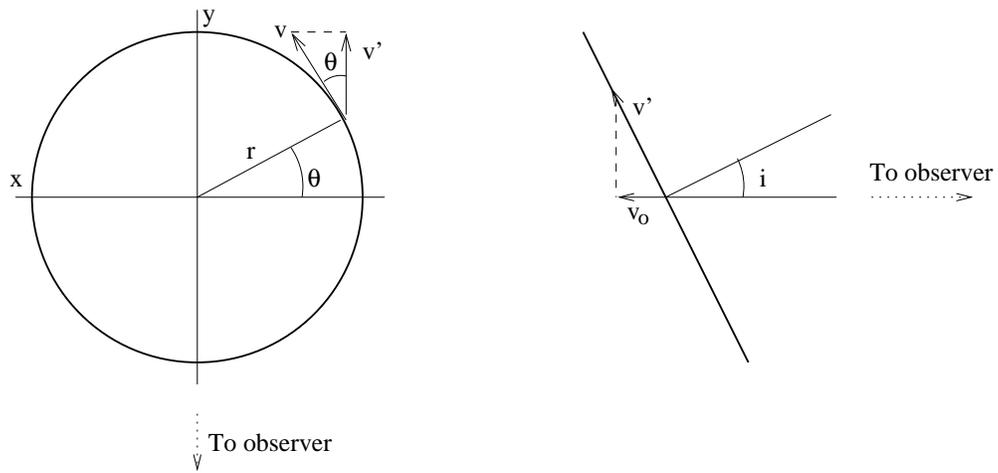}
\caption{\label{figprojected} Geometry of the disk and projections
of the Keplerian velocity of the gas, with respect to the
observer. The disk seen face-on, is shown at the left. At the right,
the disk is located edge-on, along the direction of vector $v'$. The
projected velocity along the line of sight is $v_\circ$.}
\end{figure}

\begin{figure}
\plottwo{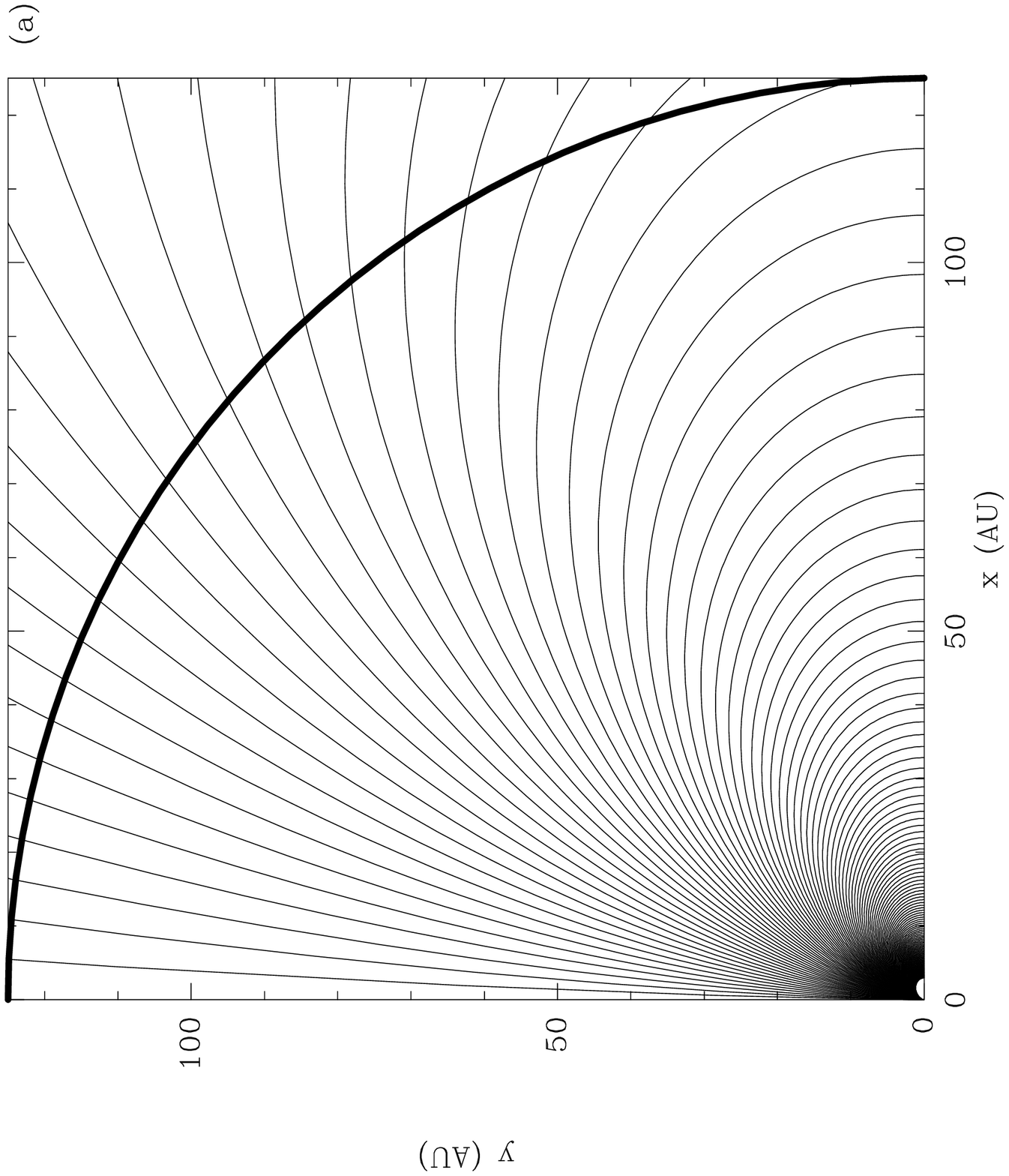}{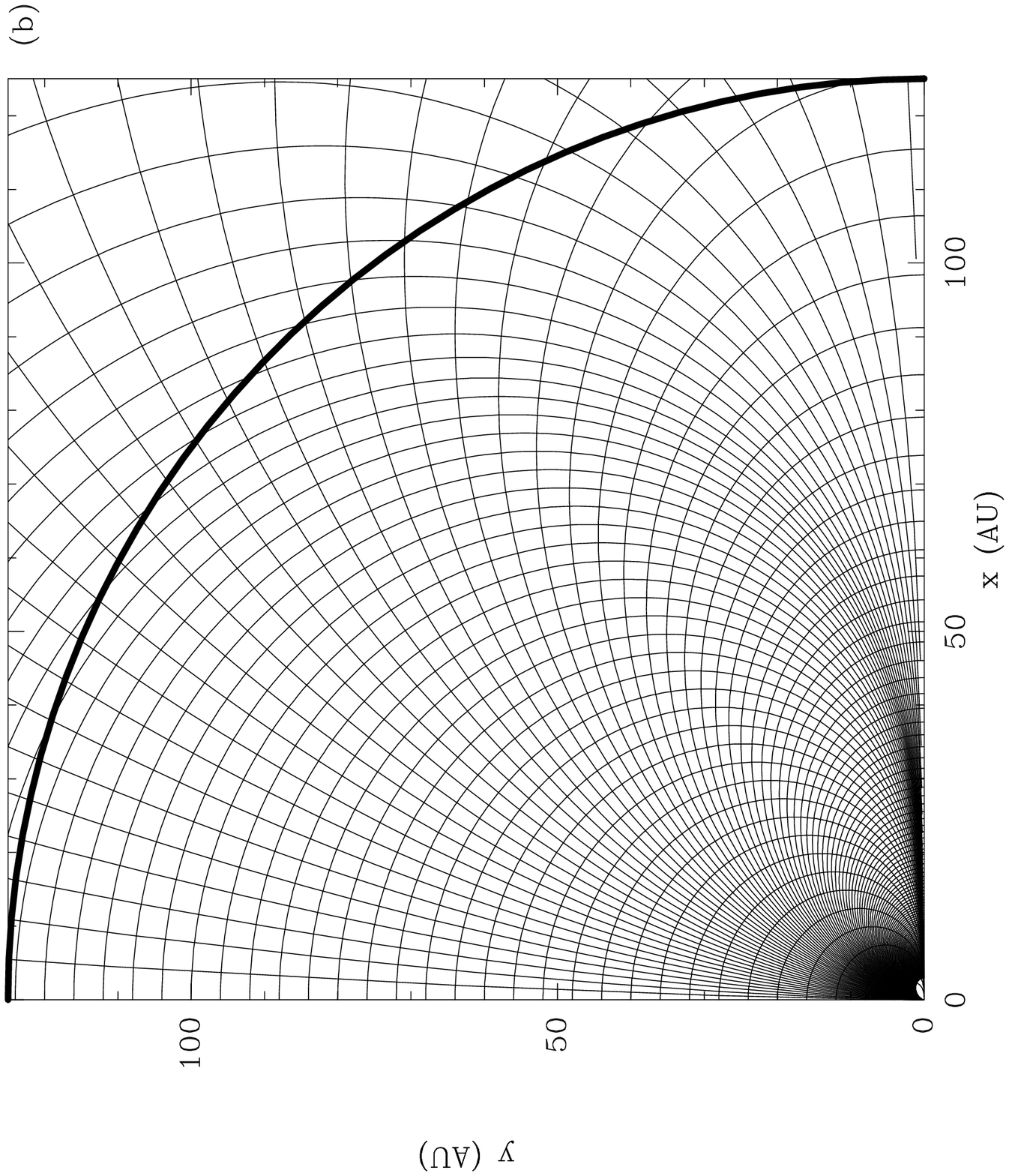}
\caption{\label{figgrid} (a)
Lines of equal projected velocities in the NW quadrant of a disk
with inclination angle
$i=60^\circ$. The interval between adjacent lines is 0.1 \kms. The
thick curve is disk outer radius. (b) Example of the type of grid used
in the integration. It is formed by the isovelocity lines shown in
(a), and their perpendicular lines.}
\end{figure}

\begin{figure}
\plotone{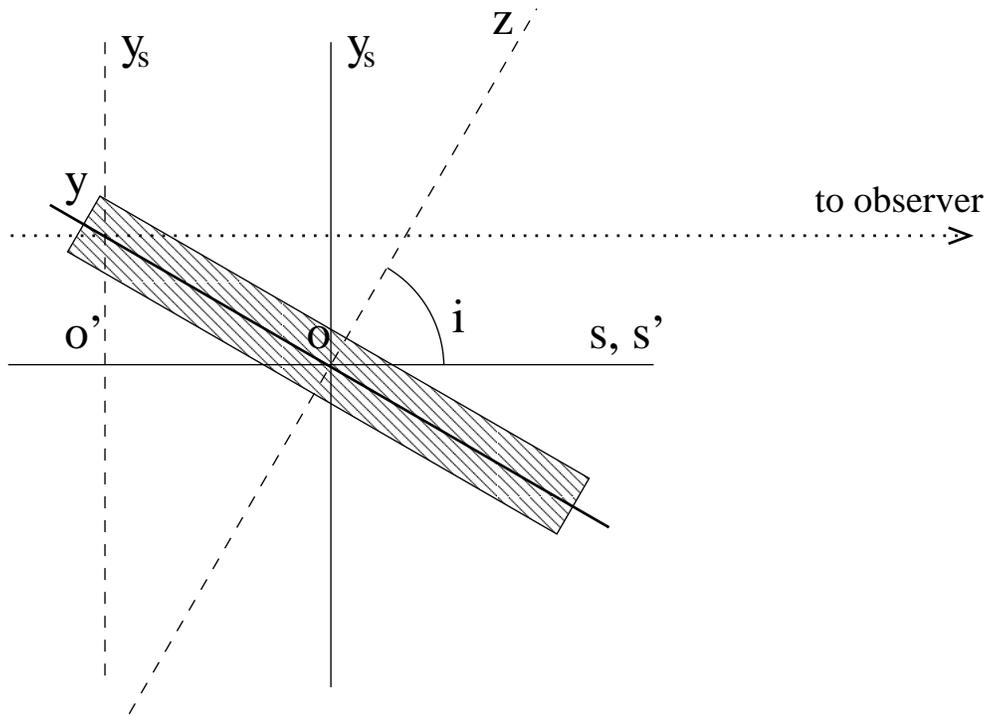}
\caption{\label{figcoordinates}
Geometry of the coordinate system used in the integration of the transfer
equation (see text).}
\end{figure}

\clearpage

\begin{deluxetable}{lccl}
\tablecaption{\label{moleculartab} Molecular Parameters}
\tablehead{
   \colhead{Molecule} & \colhead{Dipole Moment} & 
   \colhead{$X_{\rm mol}$} \tablenotemark{a} & \colhead{References}\\
   & \colhead{(Debye)} &&
}
\startdata
C$^{18}$O &$0.112$  &$2.0 \times 10^{-7}$ \tablenotemark{b}  & 1, 2   \nl
C$^{17}$O &$0.112$  &$5.0 \times 10^{-8}$  & 1, 3   \nl
CS        &$1.957$  &$2.0 \times 10^{-10}$ \tablenotemark{c} & 4, 5   \nl
C$^{34}$S &$1.957$  &$8.8 \times 10^{-12}$ \tablenotemark{d} & 4   \nl
NH$_3$    &$1.468$  &$1.0 \times 10^{-8}$  & 6, 7  \nl
\enddata
\tablerefs{(1) Rohlfs\markcite{Roh86} 1986, (2) Frerking et al.\
\markcite{fre82} 1982, (3) White \& Sandell 1995\markcite{whi95}, (4)
Winnewisser \& Cook\markcite{win68} 1968, 
(5) Blake et al.\ \markcite{bla92} 1992, (6) Townes \& Schawlow
\markcite{tow75} 1975, (7) Herbst \& Klemperer 1973\markcite{her73}}
\tablenotetext{a}{Molecular abundance relative to H$_2$}
\tablenotetext{b}{Using $X_{\rm mol}(\mbox{CO}) = 10^4$ and
terrestrial ratio C$^{17}$O/CO}
\tablenotetext{c}{Assuming lower abundance from Blake et al.\ 1992}
\tablenotetext{d}{Assuming terrestrial ratio C$^{34}$S/CS}
\end{deluxetable}

\clearpage
\begin{deluxetable}{lrccc}
  \tablecaption{\label{sensit_tab04} Line Intensities and Telescope
   Sensitivities  
   for 0\farcs4 Resolution Observations}
  \tablehead{  \colhead{Line} & \colhead{$\nu_\circ$} &
   \colhead{$F_\nu$ (0\farcs4)} & \colhead{Telescope} &
   \colhead{Sensitivity $(1\sigma)$} \\
   & \colhead{(GHz)} & \colhead{(mJy beam$^{-1}$)} && \colhead{(mJy
   beam$^{-1}$)}  }
\startdata
C$^{18}$O$(2\rightarrow 1)$ & 219.5604000 & 120 & MMA & 0.5 \nl
                                              &&& SMA & 30 \nl
C$^{18}$O$(3\rightarrow 2)$ & 392.3305722 & 220 & MMA & 0.7 \nl
                                              &&& SMA & 80 \nl
C$^{17}$O$(2\rightarrow 1)$ & 224.7143680 & 140 & MMA & 0.5 \nl
                                              &&& SMA & 30 \nl
C$^{17}$O$(3\rightarrow 2)$ & 337.0611000 & 260 & MMA & 0.7 \nl
                                              &&& SMA & 80 \nl
CS$(1\rightarrow 0)$ & 48.9909780 & 5.9 & VLA (13) & 3.0 \nl
                                      &&&VLA (27) & 1.3 \nl
CS$(2\rightarrow 1)$ & 97.9809500 & 28 & MMA & 0.4 \nl
CS$(3\rightarrow 2)$ & 146.9690330 & 58 & MMA & 0.4\nl
C$^{34}$S$(1\rightarrow 0)$ & 48.2069150 & 0.6 & VLA (27) & 1.3 \nl
C$^{34}$S$(2\rightarrow 1)$ & 96.4129400 & 7.0 & MMA & 0.4 \nl
C$^{34}$S$(3\rightarrow 2)$ & 144.6171090 & 25 & MMA & 0.4 \nl
C$^{34}$S$(5\rightarrow 4)$ & 241.0161940 & 62 & MMA & 0.5 \nl
NH$_3$(1,1) & 23.6944955 & 2.4 & VLA & 2.0 \nl
NH$_3$(2.2) & 23.7226333 & 2.5 & VLA & 2.0 \nl
NH$_3$(3.3) & 23.8701292 & 2.5 & VLA & 2.0 \nl
NH$_3$(4,4) & 24.1394163 & 2.2 & VLA & 2.0 \nl
\enddata
\tablecomments{Sensitivities calculated for 10 h of observing time
   and 1 \kms\ velocity resolution, for a disk with $i=60^\circ$. MMA
sensitivity information from Rupen\markcite{rup97} (1997), for an
array of 40$\times$8m antennas; 
SMA, from Masson et al.\markcite{mas92} (1992); VLA from its WWW page
as of 1998 March 6}
\end{deluxetable}

\clearpage
\begin{deluxetable}{lccc}
  \tablecaption{\label{sensit_tab3} Line Intensities and Telescope
   Sensitivities 
   for  $3''$ Resolution Observations}
  \tablehead{  \colhead{Line} & 
   \colhead{$F_\nu$ ($3''$)} & \colhead{Telescope} &
   \colhead{Sensitivity $(1\sigma)$} \\
   & \colhead{(mJy beam$^{-1}$)} && \colhead{(mJy
   beam$^{-1}$)}  }
\startdata
C$^{18}$O$(2\rightarrow 1)$ & 428 & OVRO & 40 \\
                                 && PdBI & 30\\
C$^{18}$O$(3\rightarrow 2)$ & 758 & SMA & 80 \\
C$^{17}$O$(2\rightarrow 1)$ & 755 & OVRO & 40 \\
                                 && PdBI & 30\\
C$^{17}$O$(3\rightarrow 2)$ & 945 & SMA & 80 \\
CS$(1\rightarrow 0)$ & 17.3 & VLA (13) & 2.7 \\
CS$(2\rightarrow 1)$ & 87 & OVRO & 30\\
                         && PdBI & 10\\
CS$(3\rightarrow 2)$ & 189 & NMA & 60\\
C$^{34}$S$(1\rightarrow 0)$ & 1.5 & VLA (13) & 2.7 \\
C$^{34}$S$(2\rightarrow 1)$ & 19 & OVRO & 30\\
                         && PdBI & 10\\
C$^{34}$S$(3\rightarrow 2)$ & 67 & NMA & 60\\
C$^{34}$S$(5\rightarrow 4)$ & 181 & SMA & 30\\
NH$_3$(1,1) & 7.7 & VLA & 2.0 \\
NH$_3$(2,2) & 7.9 & VLA & 2.0 \\
NH$_3$(3,3) & 7.8 & VLA & 2.0 \\
NH$_3$(4,4) & 4.7 & VLA & 2.0 \\
\enddata
\tablecomments{Sensitivities calculated for 10 h of observing time
   and 1 \kms\ velocity resolution, for a disk with
$i=60^\circ$. Sensitivity information has been obtained from the WWW
pages of each telescope, as of 1998 March 6}
\end{deluxetable}


\clearpage
 
\begin{references}

\reference{ada88} Adams, F. C., Shu, F. H., \& Lada, C. J.  1988, ApJ,
         326, 865
\reference{aik96} Aikawa, Y., Miyama, S. M., Nakano, T., \&
         Umebayashi, T. 1996, ApJ, 467, 684
\reference{aik97} Aikawa, Y., Umebayashi, T., Nakano, T., \& Miyama,
         S. M. 1997, ApJ, 486, 51
\reference{bal91} Balbus, S. A., \& Hawley, J. F. 1991, ApJ, 376, 214
\reference{bec89} Beckwith, S. V. W., Koresko, C. D., Sargent, A. I.
          1989, ApJ, 343, 393
\reference{bec90} Beckwith, S. V. W., Sargent, A. I., Chini, R. S., 
          \& Guesten, R. 1990, AJ, 99, 1024
\reference{bec91} Beckwith, S. V. W.,\&  Sargent, A. I. 1991, ApJ, 381, 250
\reference{bec93} Beckwith, S. V. W., \& Sargent, A. I. 1993, ApJ, 402, 280
\reference{ber88} Bertout, C., Basri, G. \& Bouvier, J. 1988, ApJ, 330, 350
\reference{bla92} Blake, G. A., Van Dishoeck, E. F., \& Sargent,
             A. I. 1992, ApJ, L99
\reference{bur96} Burrows, C. J., et al. 1996, ApJ, 473, 437
\reference{cab96} Cabrit, S., Guilloteau, S., Andre, P., Bertout, C., 
          Montmerle, T., \& Schuster, K. 1996, A\&A, 305, 527
\reference{cal91} Calvet, N., Pati\~no, A., Magris, G., \& D'Alessio, 
          P. 1991, ApJ, 380, 617
\reference{cal94} Calvet, N., Hartmann, L., Kenyon, S., \& Whitney,
          B. 1994, ApJ, 434, 330 (CHKW)
\reference{cal99} Calvet, N., Hartmann, L., \& Strom, S.E. 1999, 
 In Protostars and Planets IV, ed. V. Mannings, A. P. Boss \& S. S.
Russell (Tucson: University of Arizona Press), in press
\reference{car93} Carr, J. S., Tokunaga, A. T., Najita, J., Shu,
          F. H., \& Glassgold, A. E. 1993, ApJ, 411, 37
\reference{dal97}  D'Alessio, P., Calvet, N., \& Hartmann, L. 1997,
	ApJ, 474, 397 (DCH)
\reference{dra84} Draine, B. T., \& Lee, H. M. 1984, ApJ, 285,89
\reference{dub92} Dubrulle, B., 1992, A\&A, 266, 592
\reference{dut94} Dutrey, A., Guilloteau, S., \& Simon, M. 1994, 
          A\&A, 286, 149
\reference{fra92} Frank, J., King, A.R., \& Raine, D. J. 1992, 
          in Accretion power in Astrophysics, (Cambridge:University 
          Press), 72
\reference{fre82} Frerking, M. A., Langer, W. D., \& Wilson,
          R. W. 1982, ApJ, 262, 590
\reference{gal93a} Galli, D., \& Shu, F. H. 1993a, ApJ, 417, 220
\reference{gal93b} Galli, D., \& Shu, F. H. 1993b, ApJ, 417, 243
\reference{gom95} G\'omez, J. F., \& D'Alessio, P. 1995, in  Disks, 
	Outflows and Star Formation, ed. S. Lizano \& J. M. Torrelles,
	RevMexAASC, 1, 339
\reference{gom93} G\'omez, J. F., Torrelles, J. M., Ho, P. T. P., 
          Rodr\'\i guez, L. F., \& Cant\'o, J. 1993, ApJ, 414, 333
\reference{gra89} Grasdalen, G. L., Sloan, G., Stout, N., Strom, 
          S. E., \& Welty, A. D. 1989, ApJ, 339, 37
\reference{har94} Hartmann, L., Boss, A., Calvet, N., \& Whitney, B. 
          1994, ApJ, 430, 49
\reference{har96} Hartmann, L., Calvet, N., \& Boss, A. 1996, ApJ,
	464, 387 (HCB)
\reference{har85} Hartmann, L., \& Kenyon, S. J. 1985, ApJ, 299, 462
\reference{har87a} Hartmann, L., \& Kenyon, S. J. 1987a, ApJ, 312, 243
\reference{har87b} Hartmann, L., \& Kenyon, S. J. 1987b, ApJ, 322, 393
\reference{haw95} Hawley, J. F., Gammie, C. F., \& Balbus, S. A. 
          1995, ApJ, 440, 742
\reference{hay93} Hayashi, M., Ohashi, N., \& Miyama, S. 1993, ApJ,
          418,L71
\reference{her73} Herbst, E., \& Klemperer, W., 1973, ApJ, 185, 505
\reference{kee90} Keene, J., \& Masson, C.R. 1990, ApJ, 355, 635
\reference{ken88} Kenyon, S. J., Hartmann, L., \& Hewett, R. 1988, 
          ApJ, 325, 231
\reference{koe93} Koerner, D. W., Sargent, A. I., \& Beckwith, S. V. W. 
          1993, ApJ, 408, 93
\reference{lay94} Lay, O. P, Carlstrom, J. E., Hill, R. E., \&
          Phillps, T. G. 1994, ApJ, 434, L75
\reference{lay97} Lay, O. P., Carlstrom, J. E. \&  Hill, R. E. 
          1997, ApJ, 489, 917
\reference{mas92} Masson, C., Bloemhof, E., Blundell, R., Bruckman,
          W., Ho, P., Keto, E., Levine, M., Raffin, P., Reid, M., Wolfire,
          M. 1992, Design Study for the Submillimeter Interferometer Array of
          the Smithsonian Astrophysical Observatory
\reference{mcc96} McCaughrean, M. J., \& O'Dell, C. R., 1996, AJ, 111, 1977
\reference{mor89} Morfill, G. E. 1989, in Low Mass Star Formation and
          Pre-Main Sequence Objects, ed. B. Reipurth (Garching:ESO), 191
\reference{mun96} Mundy, L. G., Looney, L. W., Erickson, W., Grossman,
          A., Welch, W. J., Froster, J. R., Wright, M. C. H.,
          Plambeck, R. L., Lugten, J., \& Thornton, D. D. 1996, ApJ, 464, 169
\reference{muz98} Muzerolle, J., Hartmann, L., \& Calvet, N. 1998, AJ, 116, 2965
\reference{naj96} Najita, J., Carr, J. S., Glassgold, A. E., Shu,
          F. H., Tokunaga, A. T. 1996, ApJ, 462, 919
\reference{naj99} Najita, J., Edwards, S., Basri, G., \& Carr,
J.  1999, In Protostars and Planets IV, ed. V. Mannings, A. P. Boss \& S. S.
Russell (Tucson: University of Arizona Press), in press
\reference{ode93} O'Dell, C. R., Wen, Z., \& Hu, X. 1993, ApJ, 410, 696
\reference{omo92} Omodaka, T., Kitamura, Y., \& Kawazoe, E. 1992, 
          ApJ, 396, 87 
\reference{pre92} Press, W. H., Flannery, B. P., Teukolsky, S. A., \& 
          Vetterling, W. T. 1992, in Numerical Recipes in Fortran, 
          (Cambridge:University Press)
\reference{rod86} Rodr\'\i guez, L. F., Cant\'o, J., Torrelles,
          J. M., \& Ho, P. T. P. 1986, ApJ, 301, L25
\reference{rod92} Rodr\'\i guez, L. F., Cant\'o, J., Torrelles, J. M.,
          G\'omez, J. F., \& Ho, P. T. P. 1992, ApJ, 393, L29
\reference{rod94} Rodr\'\i guez, L. F., Cant\'o, J., Torrelles, J. M.,
          G\'omez, J. F., Anglada, G., \& Ho, P. T. P. 1994, ApJ, 427,
          L103
\reference{roh86} Rohlfs, K. 1986, In Tools of Radio Astronomy
          (Heidelberg:Springer) 
\reference{rup97} Rupen, M. P. 1997, MMA Memo 192
\reference{sar91} Sargent, A. I., \& Beckwith, S. V. W. 1991, ApJ, 382, 31
\reference{sha73} Shakura, N. I., \& Sunyaev, R. A. 1973, A\&A, 24, 337
\reference{sha78} Shakura, N. I., Sunyaev, R. A., \& Zilitinkevich,
          S. S. 1978, A\&A, 62, 179
\reference{sta95} Stapelfeldt, K. R., et al. 1995, ApJ, 449, 888
\reference{sta94} Stauffer, J. R., Prosser, C. F., Hartmann, L., \&
          McCaughrean, M. J. 1994, AJ, 108, 137
\reference{ter84} Terebey, S., Shu, F. H., \& Cassen, P. 1984, ApJ,
          286, 529 
\reference{tow75} Townes, C. H., \& Schawlow, A. L. 1975, In Microwave
          Spectroscopy (New York:Dover)
\reference{whi95} White, G. J., \& Sandell, G. 1995, A\&A, 299,179
\reference{wil96} Wilner, D. J., Ho, P. T. P., \& Rodr\'\i guez, L. F.
          1996, ApJ, 470, 117
\reference{win68} Winnewisser, G., \& Cook, R. L. 1968,
          J. Mol. Spect., 28, 266
\reference{zha91} Zhan, J. 1991, Ph. D. Thesis, Rice University

\end{references}
\end{document}